\documentclass[twocolumn]{aastex631} 

\newcommand\triplet{O($^{3}$P) }
\newcommand\singlet{O($^{1}$D) }

\usepackage{needspace}
\usepackage{tabularx}
\usepackage{chemfig}
\usepackage{amsmath} 
\usepackage{xcolor}

\begin{document}

\title{Methyl Isocyanate Formation from Oxygen Insertion in Methyl Cyanide Ices}

\author[0000-0002-9889-3921]{Michelle R. Brann}
\affiliation{Center for Astrophysics | Harvard $\&$ Smithsonian, 60 Garden Street, Cambridge, MA 02138, USA}

\author[0000-0001-8798-1347]{Karin I. Öberg}
\affiliation{Center for Astrophysics | Harvard $\&$ Smithsonian, 60 Garden Street, Cambridge, MA 02138, USA}

\author[0000-0003-2761-4312]{Mahesh Rajappan}
\affiliation{Center for Astrophysics | Harvard $\&$ Smithsonian, 60 Garden Street, Cambridge, MA 02138, USA}



\begin{abstract}
In cold molecular clouds, UV photolysis of icy grain mantles generates radicals that lead to new molecule formation. When radical diffusion is limited by low temperatures, oxygen atom addition and insertion reactions, enabled by photolysis of common ice components such as H$_2$O, CO$_2$, CO, and O$_3$, offer an alternative route to chemical complexity through the production of metastable, highly reactive O($^{1}D$) atoms. We examine the reactivity of these oxygen atoms generated by UV photolysis of O$_3$ with methyl cyanide (CH$_3$CN). These studies are conducted in an ultrahigh vacuum chamber at cryogenic and low-pressure conditions equipped with in situ infrared spectroscopy to monitor destruction and product formation in real time. We conclude that oxygen atoms rapidly insert into CH$_3$CN to produce primarily methyl isocyanate (CH$_3$NCO) in matrix free ices. Over the range from 10 K to 40 K, we observe no temperature dependence to either CH$_3$CN destruction or CH$_3$NCO production. When placing CH$_3$CN:O$_3$ in H$_2$O and CO$_2$ ice matrices, we find that CH$_3$NCO formation remains robust, but that the yield likely decreases due to competing reaction pathways. In the case of the H$_2$O ice we also observe a shift in product branching ratios towards alternative pathways such as the formation of hydroxyacetonitrile (HOCH$_2$CN). Overall, our results demonstrate that oxygen atom reactivity provides an important channel for generating chemical complexity from nitriles on cold grains where radical mobility is limited.

\end{abstract}

\keywords{Astrochemistry(75) --- Laboratory astrophysics(2004) --- Chemical abundances (224)}

\section{Introduction} \label{sec:intro}
Complex organic molecules (COMs) are defined as molecules with more than six atoms and at least one carbon atom \citep{herbst_complex_2009}. Among these, O-bearing and N-bearing COMs are particularly significant due to their role in prebiotic chemistry. Nitriles, organic molecules containing the \chemfig{C ~ N} (cyano) functional group, are key intermediates in the formation of biologically relevant molecules such as amino acids and RNA precursors \citep{hudson_amino_2008}. Of the almost 300 species detected in the interstellar medium (ISM), $\approx$15\% are nitriles \citep{mcguire_2021_2022}. While nitriles are often studied in the context of prebiotic Earth or Titan-like environments, they may initiate similarly rich chemistry in the ISM, where the earliest steps in molecular complexity begin. Additionally, observations of solar system comets suggest that the solar nebula also contained CH$_3$CN, along with other volatile nitriles such as hydrogen cyanide (HCN) and cyanoacetylene (HC$_3$N), at the time of comet formation \citep{altwegg_cometary_2019}.

Our work focuses on acetonitrile (methyl cyanide, CH$_3$CN) due to its presence not only in the ISM but also as the largest nitrile by the number of atoms detected in protoplanetary disks \citep{bergner_survey_2018, oberg_comet-like_2015}. In contrast to the other two nitriles seen in comets and protoplanetary disks, CH$_3$CN is likely formed and retained on icy dust grains, as its observed gas-phase abundance cannot be explained by gas-phase chemistry alone \citep{walsh_complex_2014, belloche_increased_2009, loomis_distribution_2018}. Once present on the grains, CH$_3$CN can act as a precursor to more complex species and contributes to nitrogen-bearing chemical networks through grain-surface reactions. Previous experiments determined that CH$_3$CN can isomerize or undergo UV or electron induced dissociation, producing radicals that recombine into new species such as CH$_4$, HCCCN, and CH$_3$CHNH \citep{hudson_reactions_2004, abdulgalil_laboratory_2013, bulak_photolysis_2021}. 

In addition to radical chemistry, CH$_3$CN may undergo both oxygen atom addition and insertion reactions, providing a pathway to molecular complexity. Such reactions are especially important at low temperatures where radical diffusion could be limited \citep{carder_modelling_2021}. Here, we focus \singlet atoms, the first excited electronic state of atomic oxygen as it is highly reactive, metastable, and requires no barrier to react even at cold temperatures. Previous ice experiments found that O(¹D) readily inserts into small hydrocarbons such as methane (CH$_4$) to form methanol (CH$_3$OH) and formaldehyde (H$_2$CO) \citep{bergner_methanol_2017}. The \singlet atom can react with two-carbon or unsaturated hydrocarbons to form ethanol (CH$_3$CH$_2$OH), ethylene oxide (C$_2$H$_4$O), and ketene (H$_2$CCO) \citep{bergner_oxygen_2019}. We investigate whether oxygen atom mechanisms can extend from hydrocarbons to nitriles as a method to increase chemical complexity on icy grains. 

The \singlet atom can be produced by photolysis or radiolysis of common ice mantle substituents such as H$_2$O, CO$_2$, or O$_3$ \citep{matsumi_photolysis_2003, kedzierski_production_2013, bergner_methanol_2017, bergner_oxygen_2019}. H$_2$O is the dominant component of interstellar ices and CO$_2$ is typically present at abundances of $\sim$10–30\% relative to H$_2$O \citep{boogert_observations_2015}. CO$_2$ photodissociates to O($^1$D) + CO between 120–170 nm with high quantum yields ($\sim$96\%) at 147 and 157 nm \citep{zhu_production_1990}, while H$_2$O produces O($^1$D) + H$_2$ from 105–145 nm at lower efficiency ($\sim$10\%) \citep{ung_photolysis_1974}. Additionally, both molecules generate O($^1$D) upon Ly$\alpha$ (121.6 nm) irradiation,  which is an important component of the UV field in dense cloud cores and protoplanetary disks \citep{oberg_photochemistry_2016}. We focus mainly on O($^1$D) production from O$_3$, which dissociates efficiently between 248-308 nm to O($^1$D) + O$_2$ with high quantum yield (79-94\%) without disrupting our initial reactants \citep{matsumi_photolysis_2003}. However, these alternative pathways with H$_2$O and CO$_2$ underscore the broader relevance of O($^1$D) chemistry in interstellar ices.  

We examine CH$_3$CN mixed with either O$_3$ or O$_2$ exposed to UV photolysis under cryogenic and low-pressure astrophysical conditions. In Section \ref{sec:methods} we describe the experimental chamber and \singlet atom production. In Section \ref{sec:Results}, we present our results from the CH$_3$CN:O$_3$ mixture, as well as the impact of temperature, matrix (Xe, CO$_2$, and H$_2$O), and oxygen source (O$_2$) on the destruction rate and product yield. In Section \ref{sec:Discussion}, we discuss the mechanism for our observed products as well as the astrophysical implications.

\section{Experimental Methods} \label{sec:methods}
All experiments were conducted in the ultra-high vacuum SPACECAT chamber (Surface Processing Apparatus for Chemical Experimentation to Constrain Astrophysical Theories) that was previously discussed in detail \citep{lauck_co_2015, martin-domenech_formation_2020}. This chamber has a base pressure of 10$^{-10}$ Torr and contains a Caesium Iodide (CsI) substrate that is cooled down to 10 K through a closed-cycle Advanced Research Systems (ARS) Helium cryostat and temperature regulated with a Lakeshore Model 335 controller. Our system has an accuracy of 2 K and an uncertainty of 0.1 K. 

We dose the gases through a differentially pumped gas line with a base pressure of 10$^{-4}$ hPa onto the sample substrate. The following gases are used in this study: CH$_3$CN (Millipore Sigma, 99.9 atom \%$^{14}$N), CH$_3$C$^{15}$N (Millipore Sigma, 98 \%$^{15}$N), CO$_2$ (Millipore Sigma, 99.9 atom \%$^{12}$C), deionized H$_2$O, Xe (70 \% $^{129}$Xe), and O$_2$ (Airgas, 99.99 atom \% $^{16}$O, to produce O$_3$). The sample is exposed to the UV source and ices are monitored in real time with an infrared spectrometer.  A Pfieffer QMG 220 M1 quadrupole mass spectrometer senses background gas composition in the chamber as well as desorption of ices for each experiment.

\begin{deluxetable}{lccc}
\tabletypesize{\scriptsize}
\tablewidth{0pt} 
\tablecaption{IR Peak Locations and Band Strengths}
\label{tab:band_stregnths}
\tablehead{\colhead{Species} & \colhead{Line Center} & \colhead{Band Strength} & \colhead{Reference} \\ 
\colhead{} & \colhead{(cm$^{-1}$)} & \colhead{(cm molecule $^{-1}$)} & \colhead{} } 
\startdata
CH$_3$CN & 1041 & 1.6E-18 & \citep{rachid_infrared_2022}  \\
CH$_3$CN & 920 & 0.35E-18 & \citep{rachid_infrared_2022}  \\
\hline
O$_3$ & 1038 & 8.88E-18 & \citep{loeffler_decomposition_2006} \\
CO$_2$ & 2329 & 1.10E-16 & \citep{bouilloud_bibliographic_2015} \\
H$_2$O & 3280 & 2.2E-16 &    \citep{bouilloud_bibliographic_2015} \\
\hline
CH$_3$NCO & 2278 & 1.30E-16 & \citep{mate_laboratory_2017} \\
\enddata
\tablecomments{The error on CH$_3$CN and O$_3$ is 20\%, while the error on CH$_3$NCO is 30 \%}
\end{deluxetable}

\subsection{Ice Column Densities} \label {subsec:column_densities}
All IR spectra were analyzed using single or multi-Gaussian fits following the subtraction of a local linear or cubic baseline. Spectra were acquired with a Bruker Vertex 70v Fourier transform infrared spectrometer (FTIR) in transmission mode with a liquid N$_2$ cooled mercury cadmium telluride (MCT) detector. Each infrared spectra is an average of 128 scans taken using 1 cm$^{-1}$ resolution with a clean CsI for the background reference spectra at the experimental temperature. Molecule column densities $N_i$ are calculated as 

\begin{equation}
    N_i= 2.3\frac{\int Abs(\tilde{\nu}) d\tilde{\nu}}{A_i}
\end{equation} \label{equation:ML}

where $\int Abs(\tilde{\nu}) d\tilde{\nu}$ is the integrated IR absorbance band and $A_i$ is the band strength. We convert column densities to monolayers (ML) using the relation that 1 ML is 1 $\times$ 10$^{15}$ molecules cm$^{-1}$ \citep{callen_adsorption_1990}. Table \ref{tab:band_stregnths} contains the band strengths for the species of interest. While CH$_3$CN has multiple infrared absorption bands, most overlap with product features or O$_3$. The C–C stretching mode at 920 cm$^{-1}$ is the only isolated feature and is therefore used to determine CH$_3$CN column density. 

The O$_3$ column densities previously reported in \citep{loeffler_decomposition_2006} are Reflection Absorption Infrared Spectroscopy (RAIRS) measurements. First, we assume that the ratio between RAIRS and transmission band strengths remains constant within a specific experiment and apparatus \citep{ioppolo_laboratory_2008}. To determine the transmission band strength of O$_3$, we scale the RAIRS value reported in \citet{loeffler_decomposition_2006} using the empirically determined transmission-to-RAIRS ratio for the H$_2$O $\nu_2$ bending mode from \citet{bouilloud_bibliographic_2015}, which was derived using the same apparatus.

There is blending between the CH$_3$ rock of the CH$_3$CN at 1041 cm$^{-1}$ and the O$_3$ $\nu_3$ stretch at 1038 cm$^{-1}$. In order to calculate the integrated IR absorbance band of the O$_3$ band to determine the O$_3$ column density, we first determine the CH$_3$CN column density from the C-C stretching mode. Then, using the relative band strength of the C-C stretching mode at 920 cm$^{-1}$ to the CH$_3$ rock mode at 1038 cm$^{-1}$, we subtract off the band area and use the corrected band area to calculate the O$_3$ column density. Section \ref{subsec:destruct_product} details the ice column density measurements of the identified products.

\subsection{Ozone (O$_3$) Production and Deposition} \label{subsec:ozone}
We produce O$_3$ by electrically discharging high purity O$_2$ into O$_3$ with a Nano 15 Ozone Generator (Absolute Ozone). O$_2$ is continuously flowed through the O$_3$ cell at a pressure between 15 - 20 psi. After ozone production, the pressure is stepped down to a few Torr using needle valves and a variable leak valve allowing O$_3$ to be slowly and controllably dosed into the chamber. During O$_3$ production, the output remains a mixture of O$_3$ and O$_2$. To minimize O$_2$ incorporation into the O$_3$ ice, we deposit O$_3$ at 40 K \citep{collings_laboratory_2004}. Under our experimental conditions, this results in an O$_3$ deposition rate of $\sim$3 monolayers (ML) per minute and minimal O$_2$ contamination. For all experiments, O$_3$ is co-deposited with all the other ice components (see Section \ref{subsec:experimental_details} for more details). 

\subsection{UV Lamp} \label{subsec:UV_lamp}

We employ an Analytik Jena UVP Pen-Ray Lamp (Model 11SC-1) to generate ultraviolet (UV) radiation at 254 nm. At 254 nm, UV photolysis of O$_3$ produces excited state atomic oxygen as shown in Equation \ref{equation:O3} \citep{matsumi_photolysis_2003}.
\begin{equation}
    O_3 + h\nu \  (\lambda < 310 nm) \rightarrow O(^{1}D) + O_2
\label{equation:O3}
\end{equation} 

The photon flux from the UV lamp was calculated with a NIST calibrated photodiode and found be to 1 $\times$ 10$^{13}$ photons s$^{-1}$ cm$^{-2}$ at 254 nm. Control experiments confirmed that CH$_3$CN does not absorb or dissociate at this wavelength (see Figure \ref{fig:appendix_blank} in the Appendix). 

When examining CH$_3$CN:O$_2$ in Section \ref{subsec:o2}, we instead employ a H$_2$D$_2$ lamp (Hamamatsu L11798) with an emission profile between 120 nm and 160 nm \citep{bergner_methanol_2017} since O$_2$ is photostable at 254 nm. Over this wavelength range, UV photolysis of O$_2$ produces a mixture of \singlet and \triplet atoms as shown in Equation \ref{equation:O2} \citep{lee_quantum_1977}.
\begin{equation}
    O_2 + h\nu \  (\lambda = 160 nm) \rightarrow O(^{1}D) + O(^{3}P)
\label{equation:O2}
\end{equation} 

The photon flux from this lamp is 7 $\times$ 10$^{13}$ photons s$^{-1}$ cm$^{-2}$. As expected from previous studies \citep{canta_formation_2023}, control experiments determined that this lamp can photodissociate CH$_3$CN in addition to O$_2$.

\subsection{Experimental Details} \label{subsec:experimental_details}
Table \ref{tab:exp_summary} lists all the experiments. For all mixtures, we co-deposited O$_3$ or O$_2$ with any additional components (CH$_3$CN, CO$_2$, H$_2$O, Xe) to create a mixed film. After dosing, we irradiated for 2 to 4 hours with the UVP Pen-Ray Lamp at 254 nm or the H$_2$D$_2$ Lamp at 160 nm. During irradiation, IR scans were taken every 10 to 30 minutes. After the irradiation, the sample substrate was heated at a rate of 2 K/min up to 300 K to perform temperature programmed desorption (TPD). During the TPD, IR scans were collected every 5 minutes and desorbing species were monitored with the QMS.
\footnote{All data products are available on Zenodo at \href{https://doi.org/10.5281/zenodo.15556557} {https://doi.org/10.5281/zenodo.15556557}}

\begin{deluxetable*}{c c c c c c c c c} 
\tabletypesize{\scriptsize}
\label{tab:exp_summary}
\tablewidth{0pt} 
\tablecaption{Experiment Summary}
    \tablehead{\colhead{Exp} & \colhead{Deposition Temp} & \colhead{Irradiation Temp} & \colhead{Composition} & \colhead{Ratio} & \colhead{Thickness} & \colhead{CH$_3$CN Destruction}& \colhead{CH$_3$NCO Yield}\\
\colhead{} & \colhead{(K)} & \colhead{(K)} & \colhead{} & \colhead{} & \colhead{(ML)$^{1}$} & \colhead{($\%$)$^{2}$}& \colhead{(\%)$^{3}$}} 
    \startdata
    1    &	40	&	40	&	CH$_3$CN:O$_3$	&	1:1     &	280$\pm$40   & 20$\pm$3 & 26$\pm$8\\
    2    &	40	&	10	&	CH$_3$CN:O$_3$	&	1:1	    &	60$\pm$9    & 18$\pm$3 &35$\pm$10\\
    3   &	40	&	20	&	CH$_3$CN:O$_3$	&	1:1	&	64$\pm$9  &24$\pm$4 &26$\pm$8\\
    4   &	40	&	40	&	CH$_3$CN:O$_3$	&	1:1	&	60$\pm$9	  & 22$\pm$3 & 29$\pm$9\\
    5   &	40	&	40 	&    CH$_3$CN:O$_3$:Xe &  1:0.6:5 & 	$\sim$236$^{4}$ & 14$\pm$2 &  24$\pm$8 \\
    6	&	40	&	40 	&	CH$_3$CN:O$_3$:CO$_2$	&	1:1.3:4	&	150$\pm$9&	12$\pm$2 & 4.6$\pm$2.5$^{5}$\\ 
    7   &	40	&	40 	&   CH$_3$CN:O$_3$:H$_2$O  & 1:3.9:7.5   & 450$\pm$20 & 25$\pm$4 & 2.2$\pm$1.4$^{5}$  \\
    8   &	40	&	40 	&	CH$_3$C$^{15}$N:O$_3$	&	1:0.4	& 44$\pm$7 & 16$\pm2$& 25$\pm$7\\
    9 &    10  &   10  &   CH$_3$CN:O$_2$  &  1:5 & $\sim$145$^{4}$ & 80$\pm$13 & $\sim<$4.8 \\
    \color{red}10 &    40  &   40  &   CH$_3$CN  &  n/a &  41$\pm$8 & $<$1.0$\pm$0.1 & 0 \\
    \enddata
\tablenotetext{1}{The total thickness includes error from the Gaussian fitting as well as 20\% on all band strengths.}
\vspace*{-\baselineskip}
\tablenotetext{2}{The CH$_3$CN destruction was calculated after a total fluence of 1.3 × 10$^{17}$ photons cm$^{-2}$ and includes a 15$\%$ error to account for experimental variation between experiments.}
\vspace*{-\baselineskip}
\tablenotetext{3}{The reported conversion percentage from CH$_3$CN to CH$_3$NCO incorporates a 20$\%$ error into the CH$_3$CN band strength and a 30$\%$ error into the CH$_3$NCO band strength.}
\vspace*{-\baselineskip}
\tablenotetext{4}{The total thicknesses are estimated using a 5:1 ratio of Xe (Exp 5) or O$_2$ to CH$_3$CN (Exp 9). The CH$_3$CN and O$_3$ total thickness is $57 \pm 9$ (Exp 5), and the CH$_3$CN layer has a thickness of $25 \pm 5$ (Exp 9).}
\vspace*{-\baselineskip}
\tablenotetext{5}{The lower limits of CH$_3$NCO yield are calculated using a cubic baseline (see Section \ref{subsec:addtional_ices} for full details); upper limits using a linear baseline are $14 \pm 9$ for CH$_3$CN:O$_3$:CO$_2$ and $11 \pm 3$ for CH$_3$CN:O$_3$:H$_2$O.}
\vspace*{-\baselineskip}
\end{deluxetable*}


\section{Results} \label{sec:Results}
\subsection{Reactivity and Product Identification}
\begin{figure*} 
\plotone{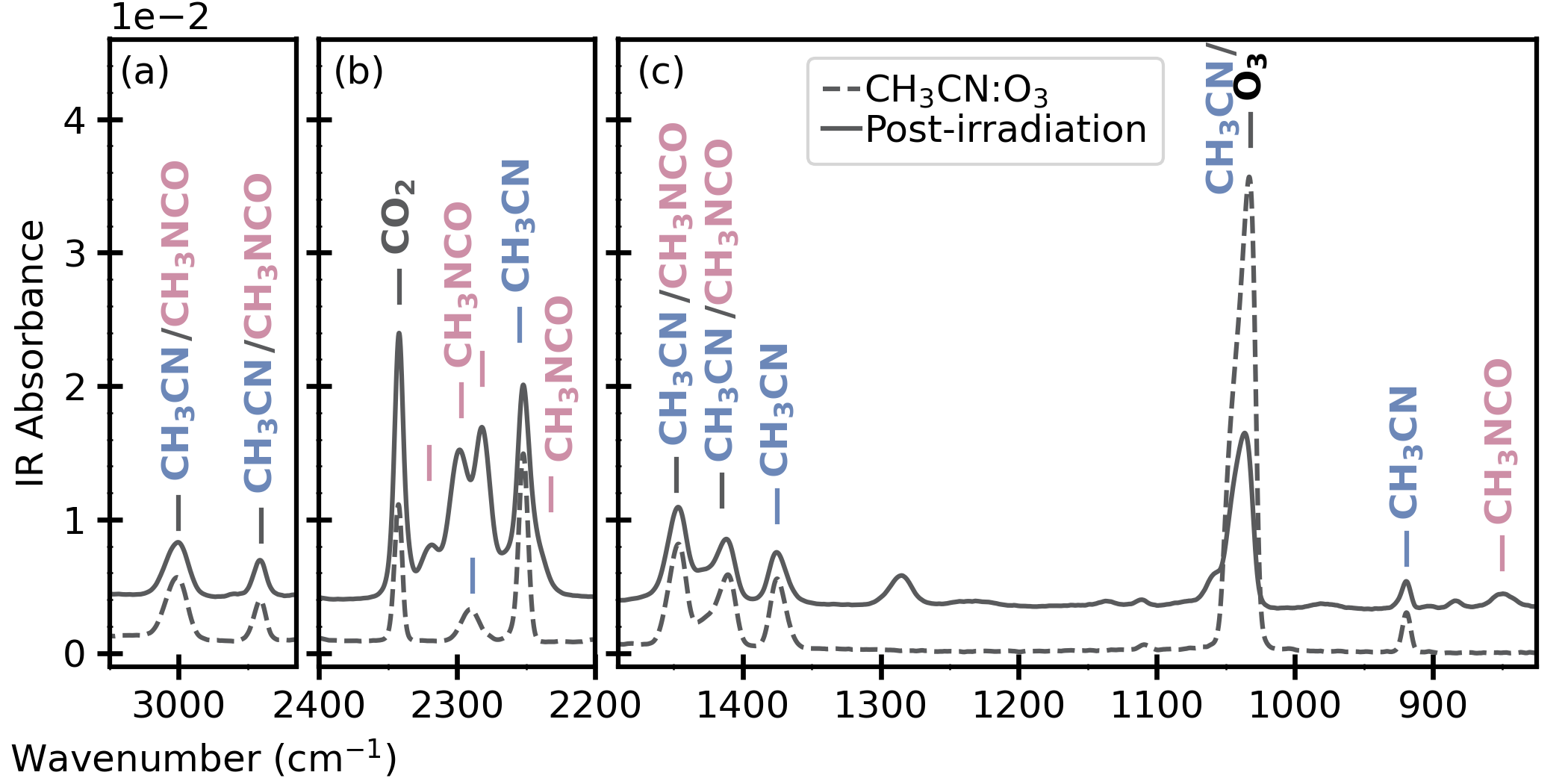}
\caption{IR spectra of characteristic regions of a 280-layer methyl cyanide and ozone (CH$_3$CN:O$_3$) ice before (dashed) and after (solid) exposure to 254 nm UV at 40 K (Experiment $\#1$). As a result of the UV irradiation, the total signal intensity decreases for O$_3$ (1037 cm$^{-1}$) and CH$_3$CN (3000, 2940, 2255, 1448, 1415, and 1375 cm$^{-1}$). There is also the growth of new peaks corresponding to methyl isocyanate (CH$_3$NCO, 2320, 2297, 2282, and 2235 cm$^{-1}$). The post-irradiated spectra is offset for clarity. 
\label{fig:before_after}}
\end{figure*}

We identify reactants and products through IR spectral analysis of the ices and QMS measurements during ice desorption. To confirm product assignments, we also conduct additional experiments with CH$_3$C$^{15}$N and analyze isotopic shifts in both the IR and QMS data. 

Figure \ref{fig:before_after} shows an IR spectra of a mixed CH$_3$CN:O$_3$ ice at 40 K before and after UV irradiation for three and a half hours at 254 nm. Prior to UV exposure, spectral features are easily correlated with gas-phase and condensed phase CH$_3$CN and O$_3$ peak assignments \citep{abdulgalil_laboratory_2013, rachid_infrared_2022, hudson_reactions_2004, loeffler_decomposition_2006}. In addition to our reactants, we also identify trace amount (less than 1\%) of CO$_2$ \citep{bouilloud_bibliographic_2015}.

There are three spectral regions of particular interest in Figure \ref{fig:before_after}. First, there is the spectral feature at 920 cm$^{-1}$ that corresponds to the C-C stretch \citep{rachid_infrared_2022, abdulgalil_laboratory_2013, dhendecourt_time_1986} (Figure \ref{fig:before_after}c). We use this feature to quantify CH$_3$CN due to its isolation from any other reactant or product peaks. Additionally, spectral features at 1447, 1423, and 1374 cm$^{-1}$ are attributed to the -CH$_3$ combination, antisymmetric, and symmetric deformation modes of CH$_3$CN \citep{rachid_infrared_2022, carvalho_photolysis_2020} (Figure \ref{fig:before_after}c).  Aside from CH$_3$CN, the feature at 1038 cm$^{-1}$ (Figure \ref{fig:before_after}c) is assigned to the $\nu_3$ O$_3$ stretch \citep{chaabouni_infrared_2000}. Second, the initial mixture (dashed) has a prominent feature at 2252 cm$^{-1}$, attributed to the CN stretch of CH$_3$CN as well as a combination of modes at 2289 cm$^{-1}$ (Figure \ref{fig:before_after}b). Third, there are additional CH$_3$ stretching modes at 3000 and 2940 cm$^{-1}$ (Figure \ref{fig:before_after}a).

\begin{figure}
\plotone{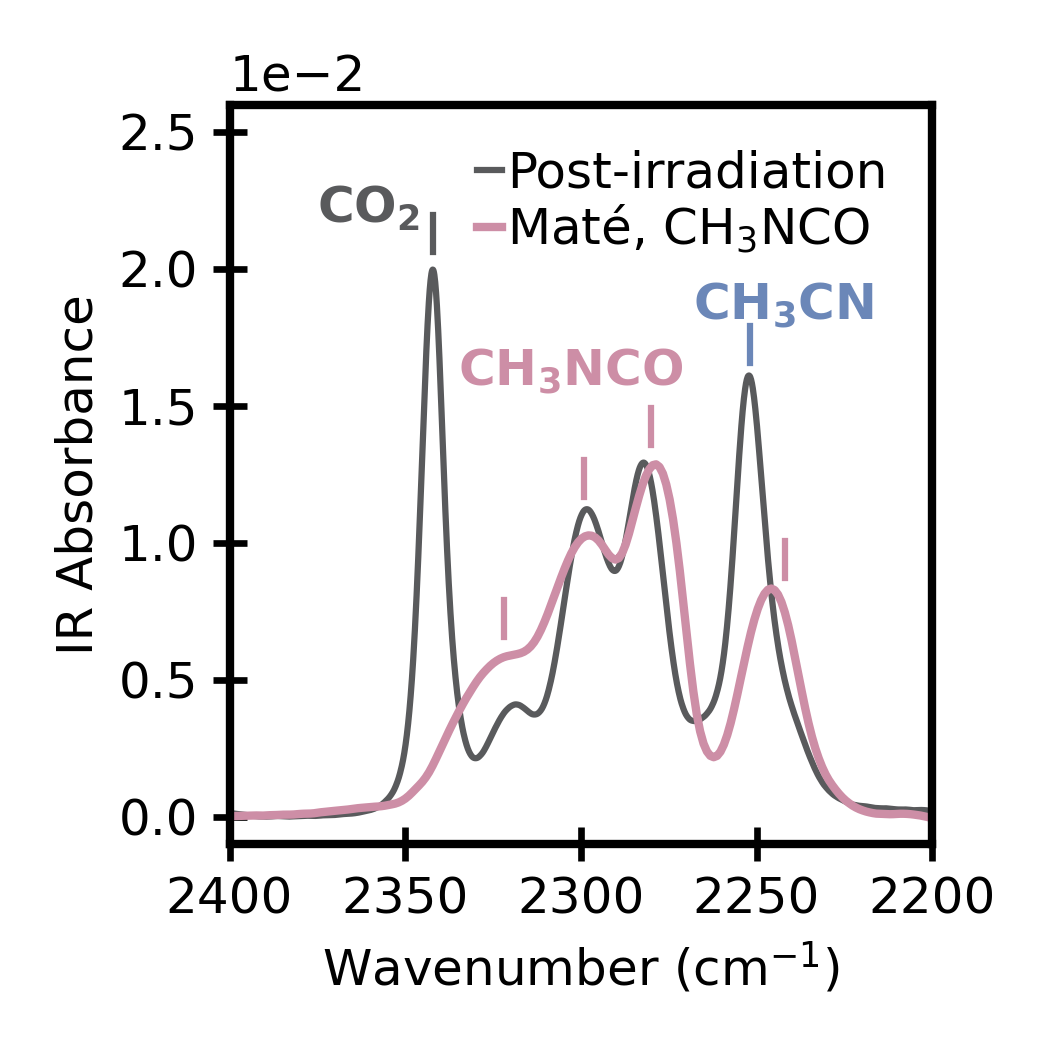}
\caption{Post-irradiation spectra overlaid with pure CH$_3$NCO spectra \citep{mate_laboratory_2017} confirms the presence of CH$_3$NCO (2320, 2297, 2282, and 2235 cm$^{-1}$). }
\label{fig:ch3cn_mate_zoom}
\end{figure}

Following  UV exposure at 254 nm, the aforementioned O$_3$ and CH$_3$CN peaks decay in intensity. The reduction in the O$_3$ peak results from UV photodissociation, while the decrease in CH$_3$CN is a secondary process initiated by reactions with atomic oxygen, as CH$_3$CN is not directly photodissociated by our lamp. At the same time, there is significant growth of novel features that represent oxygenated products. Most notably in the middle panel (Figure \ref{fig:before_after}b) the spectral signatures at 2320, 2297, 2282, and 2235 cm$^{-1}$ are assigned to the NCO asymmetric stretching mode of methyl isocyanate (CH$_3$NCO) \citep{mate_laboratory_2017, mate_stability_2018, dalbouha_structural_2016}. We further confirm our CH$_3$NCO product assignment by overlaying the post-irradiation infrared spectra with that from \citep{mate_laboratory_2017} in Figure \ref{fig:ch3cn_mate_zoom}. In addition to matching CH$_3$NCO peak locations, our post-irradiation spectra exhibit the same relative intensities of the quadruplet components as the reference spectra. Observed differences in peak widths are likely due to CH$_3$NCO being embedded in an ice matrix under our experimental conditions. Figure \ref{fig:appendix_mate} in the Appendix depicts the post-irradiation spectra overlaid with the Maté spectra \citep{mate_laboratory_2017} across the full spectral range (4000–850 cm$^{-1}$). In addition to the nitrile region between 2400 to 2000 cm$^{-1}$, there are also intense infrared features corresponding to -CH$_3$ bending and stretching modes of CH$_3$NCO in the 3000 cm$^{-1}$ and 1400 cm$^{-1}$ region (Figure \ref{fig:before_after}a, c). There is also a novel feature at 850 cm$^{-1}$ which corresponds to the CN stretch of CH$_3$NCO. 

\begin{figure} [h]
\plotone{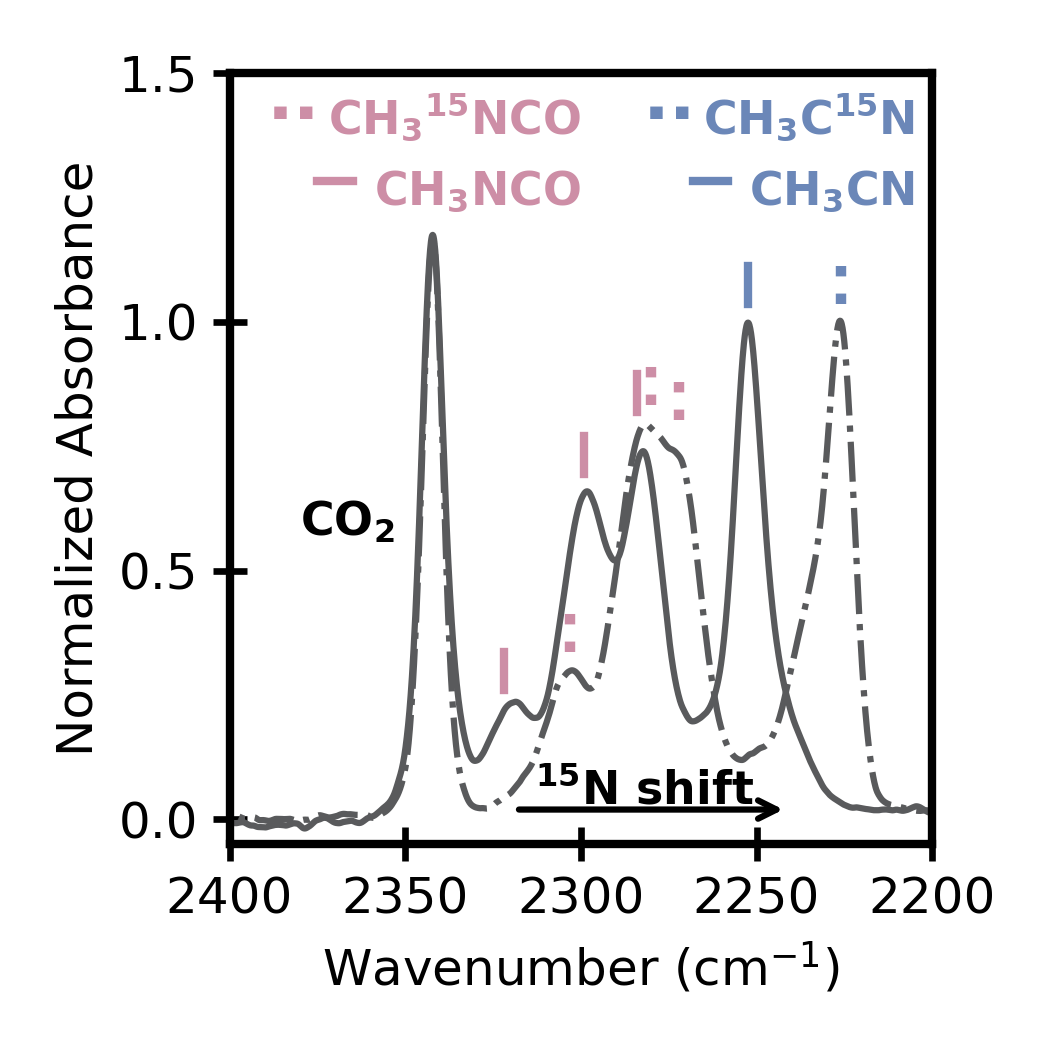}
\caption{Post-irradiation spectra of CH$_3$CN:O$_3$ (grey) and CH$_3$C$^{15}$N:O$_3$ (dashed) at 40 K. With $^{15}$N, the quadruplet NCO asymmetric stretching feature of CH$_3$NCO shifts to lower wavenumbers. Both spectra are normalized to the intensity of CN feature of CH$_3$CN (2252 cm$^{-1}$) and CH$_3$C$^{15}$N (2226 cm$^{-1}$).}
\label{fig:ch3cn_15}
\end{figure}

\begin{figure} [h]
\plotone{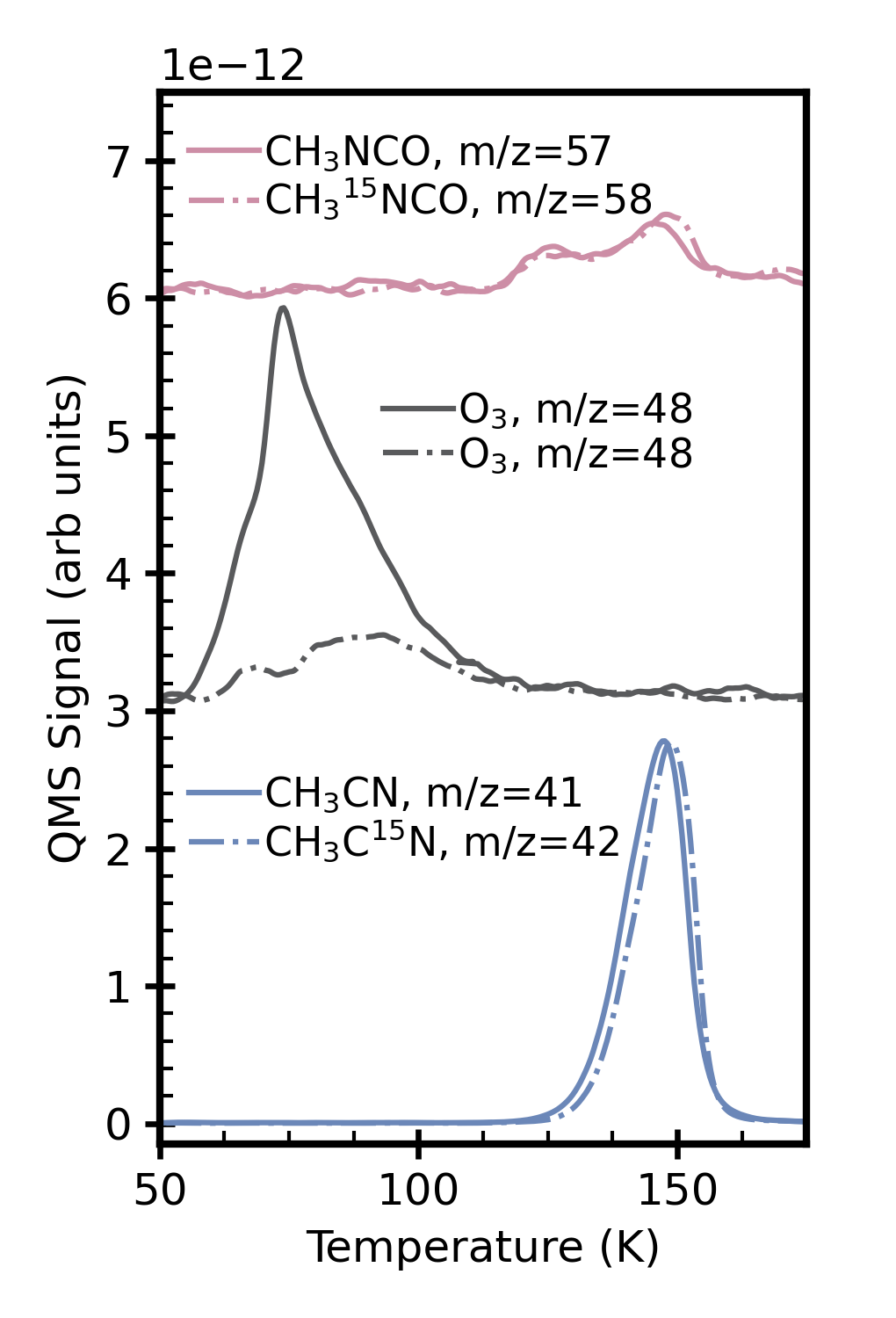}
\caption{Temperature Programmed Desorption (TPD) of CH$_3$CN:O$_3$ (solid) and CH$_3$C$^{15}$N:O$_3$ (dashed) ice after exposure to UV at 40 K confirms the presence of CH$_3$NCO (m/z = 57). There is also some O$_3$ (m/z = 48) and CH$_3$CN (m/z = 41) left on the surface. There is a 1 m/z shift for both CH$_3$C$^{15}$N (42) and CH$_3$$^{15}$NCO (58) in the $^{15}$N isotopically labeled experiment. Each experiment is normalized to the intensity of the methyl cyanide (CH$_3$CN, CH$_3$C$^{15}$N). For both experiments, the CH$_3$CN signal is 3$\%$ of the full QMS signal. 
\label{fig:mass_spec}
}
\end{figure}

To further support the identification of CH$_3$NCO as a product, we irradiated a mixture of CH$_3$C$^{15}$N:O$_3$ at 40 K. As shown in Figure \ref{fig:ch3cn_15}, we find that with $^{15}$N, the quadruplet NCO asymmetric stretching feature of CH$_3$NCO shifts to lower wavenumbers. This shift is consistent with previous studies of CH$_3$CN isotopologues (from 2252 to 2225 cm$^{-1}$) \citep{dereka_characterization_2022} and confirms CH$_3$NCO product identification. 

In addition to IR, TPD data can help confirm product identities and their relative stability on the surface. Based on mass fragmentation patterns, we assign m/z of 57 to CH$_3$NCO, m/z = 41 to CH$_3$CN, and m/z = 48 to O$_3$. As shown in Figure \ref{fig:mass_spec}, O$_3$ is the least stable on the surface with its desorption feature at 80 K \citep{sivaraman_temperature-dependent_2007}. We find CH$_3$NCO sublimation begins around 130 K, which is consistent with stability analysis in \citep{mate_laboratory_2017} and IR disappearance of the NCO asymmetric stretching mode. Finally, CH$_3$CN desorption begins around 120 K and peaks close to 150 K, consistent with CH$_3$CN multilayer binding energy studies from \citep{abdulgalil_laboratory_2013}. We observe that approximately half of CH$_3$NCO sublimates at its expected temperature (130~K), while the remainder co-desorbs with CH$_3$CN at 150 K. Since CH$_3$NCO is more volatile than CH$_3$CN, its delayed desorption suggests that a portion is entrapped within the CH$_3$CN matrix and only released upon CH$_3$CN sublimation. Such desorption behavior is consistently observed in interstellar ice analog studies \citep{bar-nun_trapping_1985, bar-nun_trapping_2007, simon_entrapment_2019}. Additionally, in the isotopic substitution experiment with $^{15}$N (dashed), there is a 1 m/z shift for both CH$_3$C$^{15}$N (42) and CH$_3$$^{15}$NCO (58) (Figure \ref{fig:mass_spec}).

From both infrared analysis of the the species on the surface and mass spec analysis of the desorbing mass fragments, we identify CH$_3$NCO as the major product for the CH$_3$CN:O$_3$ reaction. While we do not have clear evidence for any other abundant products, there are a few unidentified minor features in our infrared spectra at 1285 cm$^{-1}$, 1050 cm$^{-1}$, and 890 cm$^{-1}$. There are no obvious isotopic shifts associated with these features, which suggests that they are not nitrogen-containing products. We tentatively assign 1050 cm$^{-1}$, and 890 cm$^{-1}$ to ethanol (CH$_3$CH$_2$OH) \citep{scheltinga_infrared_2018}. Additionally, we do not identify any mass fragments above m/z = 60 in the TPD, suggesting that there are few high-molecular-weight polymeric or oligomeric species formed.

\subsection{Destruction and Product Quantification} \label{subsec:destruct_product} 

\begin{figure}
\plotone{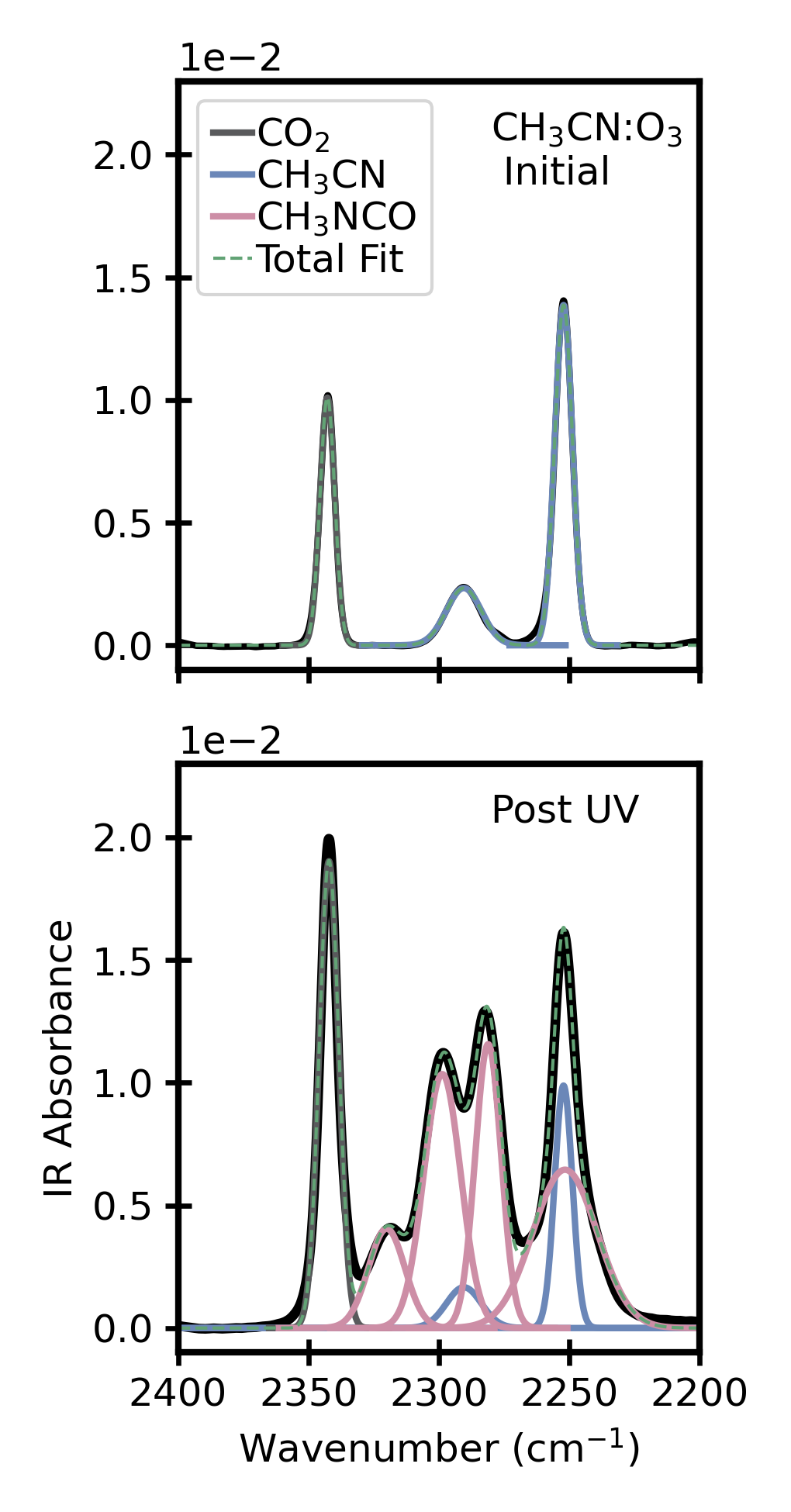}
\caption{Example fit to the assigned features in the IR spectrum of the 280 layer CH$_3$CN and O$_3$, for a representative experiment (\#1) at 40 K before (top) and after (bottom) exposure to UV at 254 nm. The measured IR spectrum is shown in black with the total fit in green. Prior to UV exposure, three Gaussians were used for the fit corresponding to CO$_2$ (gray, 2340 cm$^{-1}$), and CH$_3$CN (blue, 2289 and 2252 cm$^{-1}$). After UV exposure, we include four additional Gaussians for the growth of CH$_3$NCO (pink, 2320, 2297, 2282, and 2235 cm$^{-1}$).
\label{fig:curve_fitting_example}}
\end{figure}

In order to quantify the production of CH$_3$NCO, we consider the the nitrile region of the infrared spectra from 2400 to 2200 cm$^{-1}$ from Figure \ref{fig:before_after}. We initially select the thickest ice (experiment $\#$1) as it exhibits the greatest IR product intensity. However, our fitting procedure and resulting product yields remain consistent across all ice thicknesses. As shown in Figure \ref{fig:curve_fitting_example}, we can fit this region with 3-7 Gaussians and a linear baseline both before (top) and after (bottom) exposure to UV at 254 nm. 

\begin{figure*} [t]
\plotone{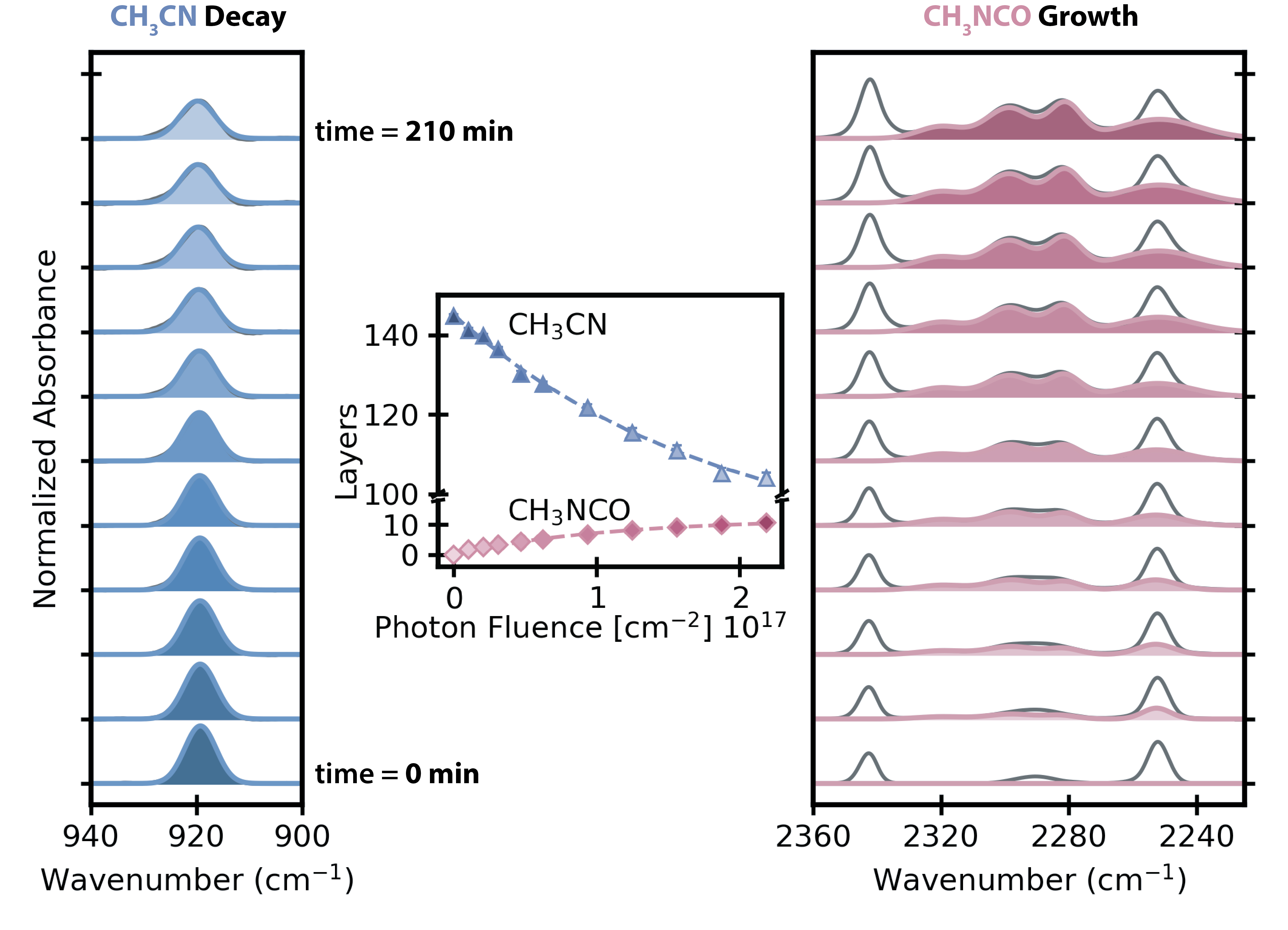}
\caption{Destruction of CH$_3$CN (left) and growth of CH$_3$NCO as a function of UV exposure for a representative experiment (\#1) at 40 K. We fit each spectrum (left) with a Gaussian curve on top and of a linear baseline and calculate the corresponding area to determine the column density and number of MLs (see Equation \ref{equation:ML}). 
\label{fig:growth_destruction}}
\end{figure*}

Due to the blending between the CH$_3$CN reactant and CH$_3$NCO product, our fitting procedure for all product IR spectra requires 3 steps. First, we use the C-C feature at 920 cm$^{-1}$ to calculate the CH$_3$CN destruction rate. In this spectral window, we fit one Gaussian on top of a linear baseline (see Figure \ref{fig:growth_destruction}). Second, we fit 3 Gaussians and a linear baseline to the IR spectra in 2400 to 2200 cm$^{-1}$ region taken prior to UV exposure. As shown in Figure \ref{fig:curve_fitting_example} top, this spectra contains only features corresponding to our initial reactants CH$_3$CN (blue, 2289 and 2252 cm$^{-1}$) as well as less than two monolayers of CO$_2$ (grey, 2340 cm$^{-1}$). Third, we fit the product region with four additional Gaussians corresponding to CH$_3$NCO growth (pink, 2320, 2297, 2282, and 2235 cm$^{-1}$) along with CH$_3$CN and CO$_2$. The heights for the CH$_3$CN features at 2289 and 2252 cm$^{-1}$ are constrained using the values derived from the destruction rate of the C-C bond (see Figure \ref{fig:growth_destruction}). Additionally, the CH$_3$CN width and peak locations are fixed from curve fitting of the initial IR spectra (Figure \ref{fig:curve_fitting_example}, top). For both the initial and post-irradiated spectra, the total fit is overlaid in green and it is in good agreement with the baseline subtracted spectra. For intermediate spectra, the width, amplitude, and peak position of each CH$_3$NCO product peak are constrained using values obtained from the post-irradiation spectrum.

For every CH$_3$CN:O$_3$ experiment, we employ this curve fitting analysis to each IR spectra recorded during UV irradiation. An example of this analysis is shown in Figure \ref{fig:growth_destruction} for both the CH$_3$CN destruction and CH$_3$NCO growth. We attribute the broadening at 920 cm$^{-1}$ to changes in ice composition as additional species form during irradiation \citep{rachid_infrared_2022}. Although seven Gaussians are fit to the IR spectral region between 2400 and 2200 cm$^{-1}$ as shown in Figure \ref{fig:curve_fitting_example} we only overlay the four CH$_3$NCO peaks for clarity. We use these four peaks to calculate the CH$_3$NCO intensity. We also calculate the destruction of O$_3$ from the integrated intensity at 1038 cm$^{-1}$  Employing the band strength and equation \ref{equation:ML}, we determine that over the three and half hour irradiation we destroy 40 layers (20\%) of CH$_3$CN, 100 layers (65\%) of O$_3$, and produce 10 layers (25\% of CH$_3$CN reacted) of CH$_3$NCO of the 279 layered CH$_3$CN and O$_3$ mixed film. 

In addition to determining the final destruction and production yields, we further quantify the kinetics of CH$_3$CN destruction and CH$_3$NCO formation by fitting the data to exponential models which assume an eventual steady state \citep{jones_mechanistical_2011}. 
\begin{equation} \label{eqn_decay}
\text{CH}_3\text{CN decay:} \quad y(x) = J + A_0 \cdot e^{-k x} 
\end{equation}
\begin{equation} \label{eqn_growth}
\text{CH}_3\text{NCO growth:} \quad y(x) = SS \cdot \left(1 - e^{-k x}\right)
\end{equation}

In Equation \ref{eqn_decay}, $y(x)$ is the CH$_3$CN signal at fluence $x$, $A_0$ is the decay magnitude, $J$ is the residual baseline absorbance remaining after decay, and $k$ is the pseudo–first-order rate constant for CH$_3$CN destruction. In Equation \ref{eqn_growth}, $y(x)$ is the CH$_3$NCO signal at a given fluence $x$, $SS$ is the steady-state absorbance achieved at long fluence, and $k$ is the pseudo–first-order rate constant for CH$_3$NCO formation. First, we find that our data is well described by these equations. Fitting yields a pseudo–first-order rate constant of $(9.6 \pm 0.8) \times 10^{-18} \, \text{cm}^2\,\text{photon}^{-1}$ for CH$_3$NCO formation, which is slightly higher than the CH$_3$CN destruction rate of $(5.4 \pm 0.8) \times 10^{-18} \, \text{cm}^2\,\text{photon}^{-1}$. Additionally, we determine that the steady state CH$_3$CN destruction yield is 60 layers, compared to 12 layers for CH$_3$NCO production, indicating that not all consumed CH$_3$CN is converted to CH$_3$NCO. Although the full carbon budget remains unaccounted, we also detect the formation of CH$_3$CH$_2$OH, as previously noted, along with the growth of CO$_2$.

\subsection{Temperature Dependence} 
\begin{figure}
\plotone{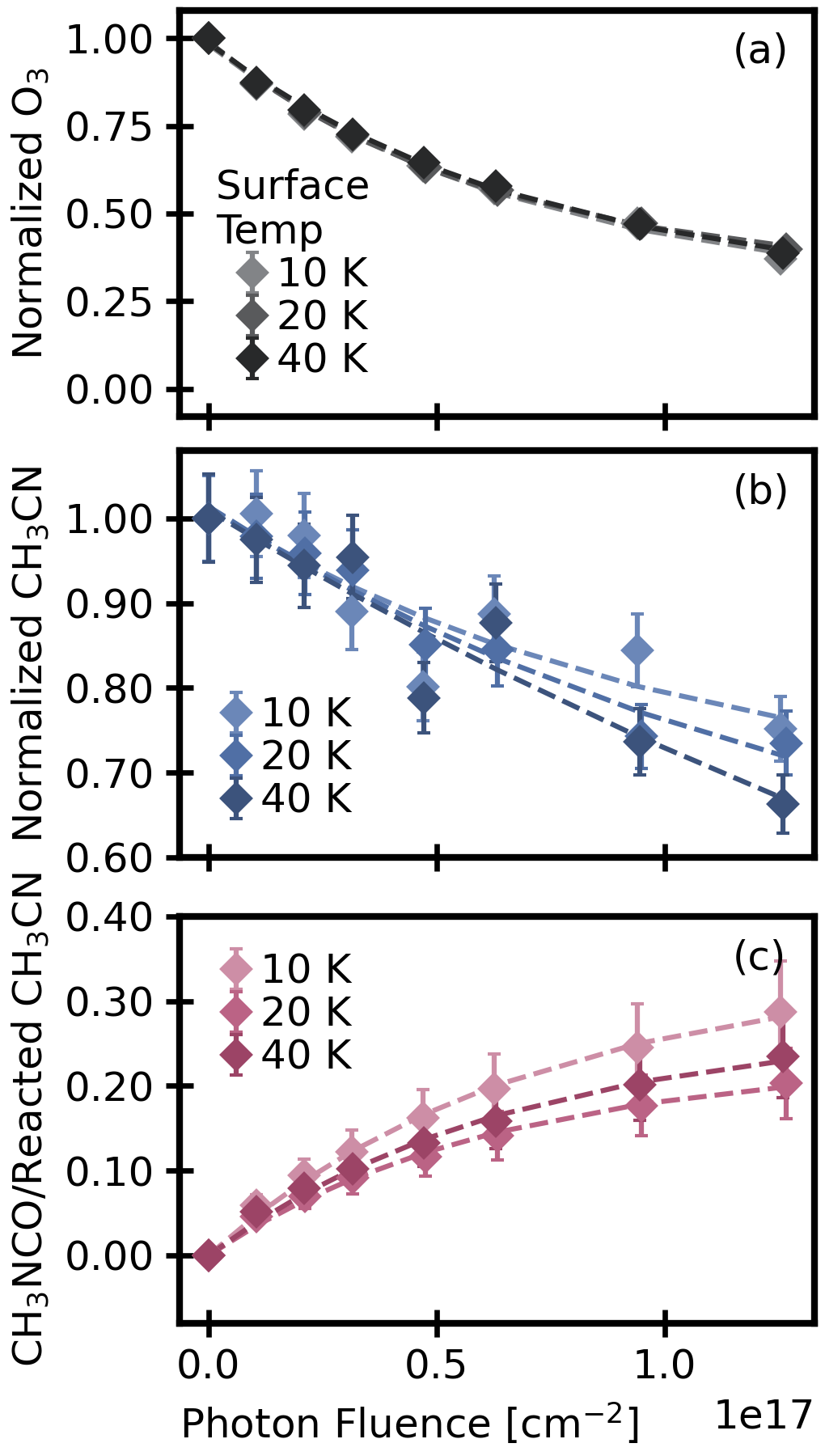}
\caption{The destruction and growth curves during CH$_3$CN:O$_3$ irradiation between 10 K and 40 K indicate that surface temperature does not impact CH$_3$NCO product formation. Destruction is calculated from integrated area of O$_3$ at 1038 cm$^{-1}$ (a) and of CH$_3$CN at 920 cm$^{-1}$ (b)  and normalized to the area prior to UV exposure. CH$_3$NCO growth is calculated from the spectral features at 2320, 2297, 2282, and 2235  cm$^{-1}$ and normalized to the amount of CH$_3$CN that reacted (c). All rates are fit to exponential decay and exponential growth functions (see Equations \ref{eqn_decay},\ref{eqn_growth}) 
\label{fig:temp_comparision}}
\end{figure}

In addition to characterizing product formation, we examine the reactivity of CH$_3$CN in 60-layer ices at 10 K, 20 K, and 40 K, containing a 1:1 ratio of CH$_3$CN to O$_3$. Each ice was deposited at 40 K prior to UV irradiation at the corresponding temperature. Due to oxygen atom desorption above 40 K \citep{collings_laboratory_2004}, we focused on the temperature range from 10 K to 40 K.  For each mixture, we exposed to UV at 254 nm for two hours while collecting infrared spectra whereby the total photon fluence was was 1.3 $\times$ 10$^{17}$ photons cm$^{-2}$. Figure \ref{fig:temp_comparision} shows the destruction of O$_3$ (a), destruction of CH$_3$CN (b), and growth of CH$_3$NCO (c). O$_3$ and CH$_3$CN are normalized to their corresponding intensity before exposure and CH$_3$NCO is normalized to the amount of CH$_3$CN that reacts. 

We observe no significant temperature dependence for the destruction of O$_3$ and CH$_3$CN between 10 K and 40 K. While CH$_3$CN has a greater scatter compared to O$_3$, the variation does not follow a systematic temperature dependence over the 10 to 40 K range. Additionally, the total destruction of O$_3$ is greater than that for CH$_3$CN. We also see no evidence for a temperature dependent CH$_3$NCO formation between 10 K and 40 K. The lack of temperature dependence on the reaction rate is consistent with gas phase \citep{hickson_kinetic_2022} and theoretical studies \citep{sun_theoretical_2010}, which explained this result by stating that \singlet atom insertion into CH$_3$CN was barrierless.

We use the same destruction and growth models (Equations \ref{eqn_decay} and \ref{eqn_growth}) to quantify CH$_3$CN loss and CH$_3$NCO formation, and apply the destruction model (Equation \ref{eqn_decay}) to O$_3$. From this analysis, we find no significant difference in the behavior of CH$_3$CN and CH$_3$NCO compared to the thicker film (Experiment$\#$1). Notably the destruction cross section of O$_3$ at 40 K is $(1.4 \pm 0.1) \times 10^{-17} \text{cm}^2\,\text{photon}^{-1}$, an order of magnitude higher than that of CH$_3$CN. Furthermore, for both O$_3$ and CH$_3$CN, the cross sections and final experimental yields are consistent with one another, supporting a coherent quantitative picture of the chemistry.  This suggests that only a portion of the \singlet atoms produced go on to react with CH$_3$CN. We determine that one of every four molecules of CH$_3$CN destroyed becomes CH$_3$NCO (Figure \ref{fig:temp_comparision}c). This conversion yield is identical to that calculated for our thickest experiment ($\#1$), supporting the conclusion that up to a $\sim$300 layered ice, thickness does not impact the CH$_3$NCO production rate. Table \ref{tab:appendix_constants} in the appendix lists all fit parameters.

\subsection{Reactivity in Interstellar Ice Analogs} \label{subsec:addtional_ices}
 
\begin{figure*} [t]
\plotone{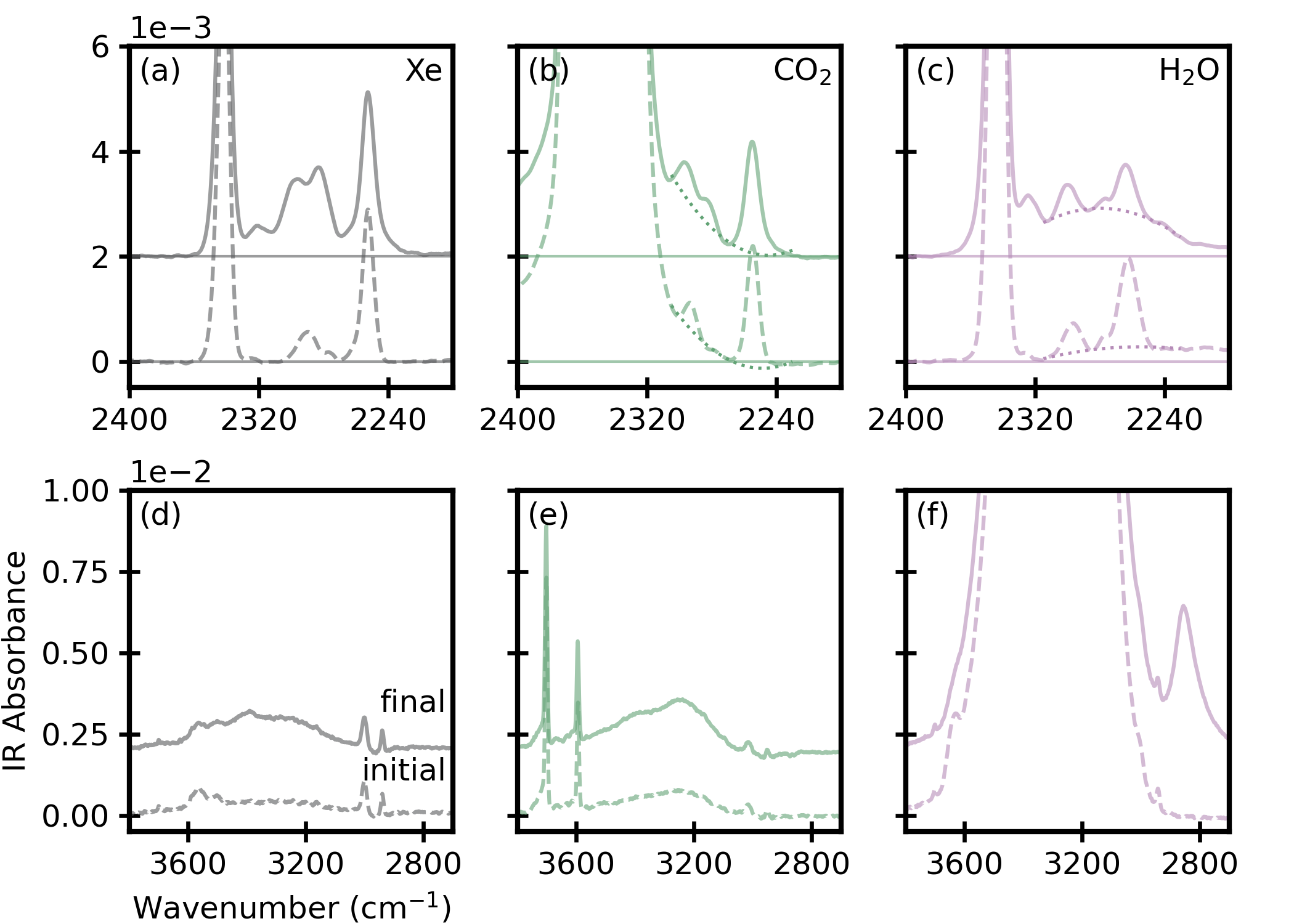}
\caption{Initial (dashed) and post-irradiation (solid) infrared spectra of CH$_3$CN:O$_3$ in a Xe (a, d), CO$_2$ (b, e), and H$_2$O (c,f) matrix at 40 K. The post-irradiated spectra is offset for clarity. The cubic (dotted) baseline for CO$_2$ and H$_2$O ices is used to calculate a lower limit on the CH$_3$NCO yield, while the linear (solid) baseline provides an upper limit.
\label{fig:matrices}}
\end{figure*}

\begin{figure}
\plotone{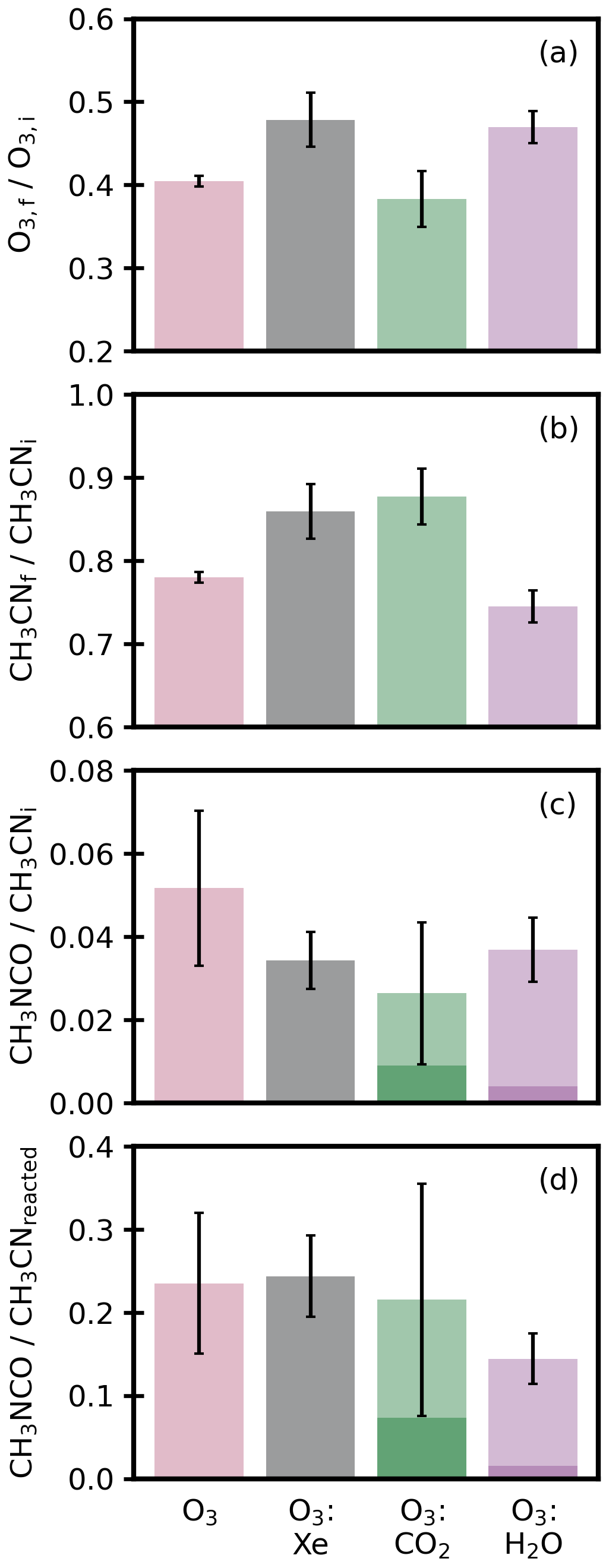}
\caption{Summary of O$_3$ destruction (a), CH$_3$CN destruction (b),  and CH$_3$NCO growth (c-d) for CH$_3$CN:O$_3$ and CH$_3$CN:O$_3$ in a Xe, CO$_2$, and H$_2$O matrices at 40 K. The CH$_3$NCO growth is normalized to the initial column density of CH$_3$CN (b) as well as to the amount of CH$_3$CN that reacted. CO$_2$ and H$_2$O matrices were fit with a cubic baseline (solid) and linear baseline to determine lower and upper limits for the CH$_3$NCO production.  CH$_3$NCO and CH$_3$CN column densities include 30\% and 20\% uncertainties, respectively, from their band strengths \citep{mate_laboratory_2017, rachid_infrared_2022}, in addition to the error from Gaussian curve fitting.
\label{fig:matrices_bar_chart}}
\end{figure}

We also consider oxygen insertion into CH$_3$CN in a variety of ice matrices that better simulate astrophysical ices. CO$_2$ and H$_2$O were selected because they are the most abundant ISM ice components apart from CO, which is too volatile to co-deposit with O$_3$ at 40 K \citep{boogert_observations_2015}.  To further constrain the reaction mechanism, we also investigate an inert Xe matrix. For each matrix, we premix CH$_3$CN with the corresponding gas to ensure uniformity throughout the ice. We explore the formation of CH$_3$NCO as well as  the emergence of any additional products. The complete IR spectra before and after exposure to UV for these ternary ices is in the appendix as Figure \ref{fig:appendix_matrices}. 

First, we examine CH$_3$CN:O$_3$ chemistry in an inert matrix of Xe (Figure \ref{fig:matrices}a) before and after 254 nm UV irradiation at 40 K. Since Xe is infrared inactive and exceeds our mass spectrometer limit (100 m/z), we estimate its abundance indirectly from the dosing rate. The reported ice thickness is thus only accurate within a factor of a few, but we estimate it to be five times greater than that of CH$_3$CN. Our CH$_3$CN column density is identical to that from the binary mixture of just CH$_3$CN:O$_3$ (experiment $\#4$). Further support for CH$_3$CN embedded in the Xe matrix is the observed splitting of CH$_3$CN combination modes at 2290~cm$^{-1}$ into peaks at 2280 and 3000 cm$^{-1}$, along with a sharpening of the nitrile mode, as expected for more isolated CH$_3$CN molecules \citep{kim_fourier_1992, knoezinger_matrix_1993}. Upon UV irradiation, the CH$_3$NCO yield relative to CH$_3$CN destruction is similar to the binary mixture (Figure \ref{fig:matrices_bar_chart}d), suggesting that formation requires only one CH$_3$CN and one O(\textsuperscript{1}D) atom.

Figure \ref{fig:matrices}b shows the reactivity of CH$_3$CN:O$_3$ in a CO$_2$ matrix at 40 K. This ice contained 150 total layers with 25 layers CH$_3$CN, 31 layers O$_3$, and 94 layers CO$_2$ to give a rough 1:1.25:4 ratio.  Before UV exposure, the IR spectra matches the binary mixture aside from the CO$_2$ feature at 2340 cm$^{-1}$. After UV irradiation at 40 K, we observe growth of CH$_3$NCO at 2297, 2282, and 2235 cm$^{-1}$. The fourth feature of the NCO band at 2320 cm$^{-1}$ is obscured by the strong carbon dioxide feature. In Figure \ref{fig:appendix_matrices}b in addition to CH$_3$NCO, there is also the growth of additional carbon and oxygenated products that we identify as carbon monoxide, CO, at 2140 cm$^{-1}$ and carbon trioxide, CO$_3$ at 1890 cm$^{-1}$ \citep{moll_carbon_1966, bouilloud_bibliographic_2015, mifsud_mid-ir_2022}. 

Figure \ref{fig:matrices}c depicts the reactivity of CH$_3$CN:O$_3$ in a H$_2$O matrix at 40 K. Here, we have 279 total layers with 24 layers CH$_3$CN, 82 layers O$_3$, and 174 layers H$_2$O to give a 1:3.9:7.5 ratio. Due to the lower IR intensity at 2297 cm$^{-1}$, CH$_3$NCO production is likely diminished compared to the other matrices. We tentatively assign the strong 2850 cm$^{-1}$ (Figure \ref{fig:matrices}f) feature to HOCH$_2$CN \citep{danger_hydroxyacetonitrile_2012, mielke_matrix_1989}, which is not observed in CO$_2$ or Xe ices (Figure \ref{fig:matrices}d, e).

To compare conversion yields across matrices, we quantified the column density of O$_3$ lost, CH$_3$CN lost, and CH$_3$NCO gained for each matrix from an IR spectra after a photon fluence of $1.3 \times 10^{17} \text{cm}^2$ (Figure \ref{fig:matrices_bar_chart}). We also include the binary CH$_3$CN:O$_3$ mixture at 40 K (pink). As shown in Figure \ref{fig:matrices_bar_chart}a and b, 52–59$\%$ of the initial O$_3$ and 12–22$\%$ of the CH$_3$CN is lost across all matrices. Notably, CH$_3$CN consumption is lowest in the CO$_2$ matrix, while O$_3$ loss is highest suggesting that oxygen atoms also react with CO$_2$, consistent with the observed 6$\%$ loss of the CO$_2$ matrix. The greater CH$_3$CN destruction in the H$_2$O matrix may reflect the higher initial O$_3$ concentration relative to CH$_3$CN in the deposited ice.

When evaluating CH$_3$NCO formation, we consider the yield both relative to the initial column density of CH$_3$CN (c) and relative to the amount of CH$_3$CN that reacted (d). First, we use the same Gaussian method with a linear baseline (Figure~\ref{fig:curve_fitting_example}). To address potential overestimation since the IR features appear to grow above the baseline we also calculate a lower limit for CH$_3$NCO production using a cubic baseline for the CO$_2$ and H$_2$O ices (dark). All baselines are annotated in Figure \ref{fig:matrices}. Table \ref{tab:exp_summary} lists CH$_3$NCO yield relative to the amount of CH$_3$CN reacted.  

As reported earlier, in the CH$_3$CN:O$_3$ mixture, approximately one out of every four CH$_3$CN molecules consumed is converted to CH$_3$NCO (Figure \ref{fig:matrices_bar_chart}d). Although total CH$_3$NCO production is slightly lower in the Xe matrix (Figure \ref{fig:matrices_bar_chart}c), the conversion efficiency remains comparable (Figure \ref{fig:matrices_bar_chart}d), indicating that that Xe matrix does not significantly alter the reaction. In contrast, incorporation of CO$_2$ and H$_2$O appears to lower the CH$_3$NCO yield up to a factor of 3 and 15 respectively. However, these values need confirmation from future studies due to the uncertainty of assigning baselines from single experiments. The lowest yield is observed in the H$_2$O matrix, consistent with the formation of alternative products such HOCH$_2$CN.  Table \ref{tab:exp_summary} summarizes the CH$_3$CN destruction and CH$_3$NCO yield for these ices, as well as for all other ices in this study. For both the CO$_2$ and H$_2$O ices, we report the conservative lower limit from the cubic baseline in the main table and the upper limit from the linear baseline as a footnote. These findings suggest that qualitatively CH$_3$NCO formation is robust under astrophysical conditions, but that the quantitative yield may change by up to an order of magnitude due to matrix composition and competing reaction pathways.

\Needspace{5\baselineskip}
\subsection{\singlet Atoms from O$_2$} \label{subsec:o2}

\begin{figure*}
\plotone{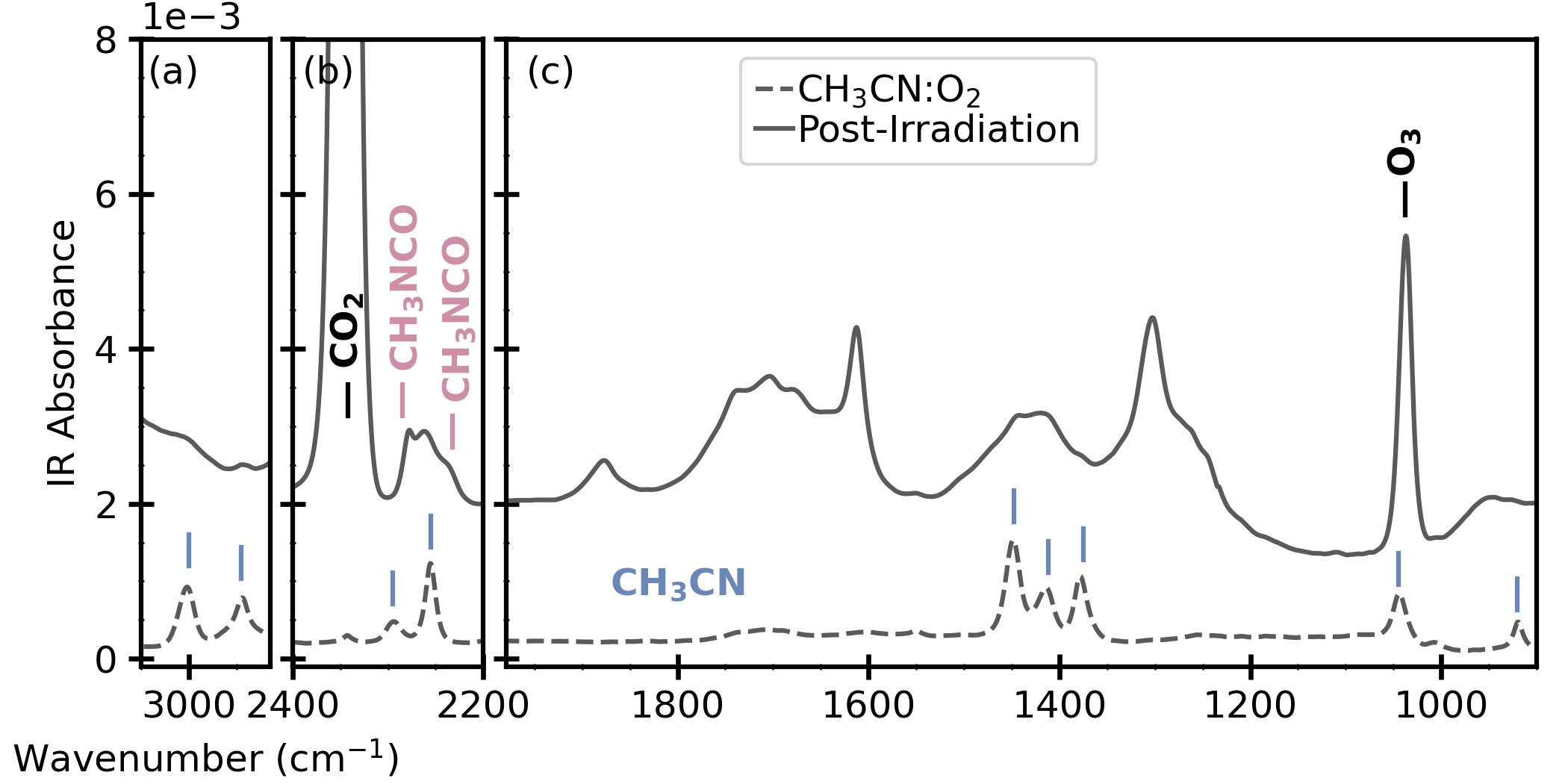}
\caption{Initial (dashed) and post-irradiation (solid) infrared spectra of CH$_3$CN:O$_2$ exposed to the H$_2$D$_2$ lamp at 10 K. As a result of UV irradiation, there are many new features, including CH$_3$NCO, CO$_2$, and O$_3$. The post-irradiated spectra is offset for clarity. 
\label{fig:ch3cn_o2}}
\end{figure*}
We also conducted experiments with CH$_3$CN and O$_2$ to investigate CH$_3$NCO formation from alternative oxygen sources. Cometary observations indicate that O$_2$ could be an important reservoir of reactive oxygen in interstellar ices \citep{altwegg_organics_2017}. Figure \ref{fig:ch3cn_o2} shows the IR spectra of the CH$_3$CN:O$_2$ ice before and after UV irradiation at 10 K using a H$_2$D$_2$ lamp. Prior to UV exposure, the spectral CH$_3$CN features parallel those in the CH$_3$CN:O$_3$ ice (see Figure 1). Upon exposure to UV between 140 nm and 175 from the H$_2$D$_2$ lamp, O$_2$ dissociates into a mixture of \singlet and \triplet atoms (Equation \ref{equation:O2}).

Since the H$_2$D$_2$ UV lamp can also dissociate CH$_3$CN, we first irradiated a pure CH$_3$CN ice to explore product formation in the absence of oyxgen atoms. We found that irradiation with the H$_2$D$_2$ lamp leads to dissociation into a variety of products, including CH$_4$, CH$_3$NC, and HCN. When examining the post-irradiation spectra of the CH$_3$CN:O$_2$ ice in Figure \ref{fig:ch3cn_o2}, we observe these same products, but also oxygen-containing products such as CO$_2$ and O$_3$, as well as CH$_3$NCO, but at a lower yield. We use the CH$_3$CN:O$_3$ experiments to identify CH$_3$NCO. These products likely result from a combination of radical chemistry and oxygen atom insertion. The CH$_3$NCO yield is quantified by summing the integrated area between 2400 and 2200 cm$^{-1}$ after a photon fluence of $1.3 \times 10^{17}$ cm$^{-2}$ and subtracting the CH$_3$CN contribution present before irradiation. Since it is possible that there are other nitrile products within this region, we consider the yield of 4.8 as an upper limit. Regardless, we find that CH$_3$NCO formation is not exclusive to O$_3$ irradiated with a 254 nm lamp and can proceed in the presence of multiple oxygen atom sources.

\section{Discussion} \label{sec:Discussion}

\subsection{CH$_3$NCO Formation Mechanism} \label{subsec:mechanism}
The only plausible reaction pathway for CH$_3$NCO formation in CH$_3$CN:O$_3$ ices involves photoproduced O atoms. Under our experimental conditions, both O($^1$D) and O($^3$P) atoms could destroy CH$_3$CN and subsequently produce CH$_3$NCO since O($^1$D) atoms produced via O$_3$ photodissociation at 254 nm \citep{matsumi_photolysis_2003} can decay to ground-state O($^3$P) on short timescales relative to the duration of the experiment. Prior studies of hydrocarbon matrices demonstrated that both \singlet and \triplet atoms can insert into unsaturated bonds, including double and triple bonds \citep{bergner_oxygen_2019, brann_reaction_2020, vanuzzo_reaction_2016, daniely_photochemical_2025}, suggesting that both spin states react under cryogenic conditions.

Despite this, several lines of evidence point to CH$_3$CN being primarily attacked by \singlet atoms in our system. First, at 254 nm, O$_3$ photodissociation yields exclusively \singlet oxygen atoms \citep{matsumi_photolysis_2003}. Second, over the range of 10 K and 40 K, we observe no temperature dependence, which suggests that the reaction proceeds without an activation barrier for either CH$_3$CN destruction or CH$_3$NCO production. This interpretation is supported by theoretical Gaussian calculations showing that while O($^3$P) reactions with CH$_3$CN typically involve a barrier \citep{sun_theoretical_2010}, the O($^1$D) pathway is predicted to be barrierless. Additional support comes from kinetic measurements in a continuous supersonic flow reactor, which found that the rate constant for O($^1$D) + CH$_3$CN is comparable to those of other known barrierless systems, such as O($^1$D) with CF$_3$CN and CH$_3$Cl \citep{hickson_kinetic_2022}.  

Additionally, \triplet oxygen atoms can lead to different reaction pathways and products than \singlet atoms. For example, matrix-isolated studies of O($^3$P) and CH$_3$CN in argon at 15 K found that although \triplet can react with CH$_3$CN, hydrogen abstraction dominates, leading to HOCH$_2$CN formation \citep{mielke_matrix_1989}. Aside from our ices containing CH$_3$CN, O$_3$, and H$_2$O, we observe no IR spectral evidence for HOCH$_2$CN, further supporting that CH$_3$CN is primarily attacked by O($^1$D) under our conditions. We note that additional experiments employing \triplet atoms as well as theoretical molecular dynamic simulations are necessary to fully confirm and examine the difference in chemistry between \triplet and \singlet atom with nitriles.

Given that CH$_3$CN is mainly destroyed by reactions with \singlet atoms, we next turn to the possible reaction pathways and expected product branching ratios. Gaussian calculations of CH$_3$CN and O atoms determined that a reaction with O($^1$D) can occur by three pathways, (1) oxygen atom insertion into the C-H bond, (2) oxygen atom insertion into the C-C bond or (3) oxygen atom addition to the carbon of the CN group  \citep{sun_theoretical_2010}. Although all three pathways are barrierless, oxygen addition to the CN group is the lowest energy state compared to the insertion into the C-H or C-C bonds. We have no evidence for oxygen insertion into the C–H bond that would produce cyanomethyl (CH$_2$CN) via the highest-energy intermediate.  Thus, we propose that for CH$_3$CN:O$_3$ ice and our conditions the reaction begins with either oxygen addition to the carbon of the CN group forming the singlet intermediate CH$_3$C(O)N or into the C-C bond forming the singlet intermediate CH$_3$OCN \citep{sun_theoretical_2010}. The CH$_3$C(O)N adopts a cyclic CNO structure with a longer CO bond and with shorter CN and CC bonds, and an ON bond length of 1.774  \AA. 

For both pathways (2) and (3), we predict either isomerization or immediate rearrangement into the more stable isocyanate structure.  We do not have IR spectral evidence for either CH$_3$NCO isomers methyl fulminate (CH$_3$ONC) or acetonitrile N-oxide (CH$_3$CNO). Thus, infrared spectra support our prediction that isocyanate (-NCO) is favored over cyanate (-OCN) and nitrile oxide (-CNO) due to its increased thermodynamic stability \citep{pasinszki_gas-phase_2001, bondybey_cno_1981}.
While we do not observe CH$_3$CN isomerization at 254 nm, its occurrence at 160 nm \citep{hudson_amino_2008} suggests that isomerization may still represent a possible pathway to CH$_3$NCO, particularly if the oxygen addition intermediate rearranges into the final, stable product.

Even when accounting for baseline and peak fitting uncertainties, we likely find a decrease in the CH$_3$NCO yield for the CO$_2$ and H$_2$O ice matrices compared to the binary CH$_3$CN:O$_3$ ice. We attribute this reduction to competing pathways but expect the same mechanism for CH$_3$NCO production, as no additional intermediate products are observed. Interestingly, in the presence of H$_2$O, we observe a new product HOCH$_2$CN. One possibility is that O(\textsuperscript{1}D) atoms react with H$_2$O to generate OH radicals, which divert reactivity toward alternate products. Additionally, H$_2$O could provide stability after oxygen insertion that enables HOCH$_2$CN formation \citep{marshall_kinetics_2024}. 

In sum, our results highlight a plausible mechanism, but transition state theory or molecular dynamics simulations are needed to confirm its viability \citep{hickson_kinetic_2022}. Our spectral identification and temperature studies in conjunction with previous theoretical studies support \singlet attack to CH$_3$CN as a barrierless route to produce CH$_3$NCO in ices.

\subsection{Astrophysical Implications} \label{subsec:implications}
We investigate CH$_3$CN reactivity with various sources of excited oxygen atoms produced from UV irradiation at cryogenic temperatures. We conclude that CH$_3$CN readily reacts with \singlet atoms generated by photodestruction of O$_3$ at 254~nm, producing CH$_3$NCO as the major oxygen and nitrogen containing product. Expanding to interstellar ice analogs, we find that the conversion yield for CH$_3$NCO is matrix dependent and we tentatively propose that less CH$_3$NCO is formed in the non-inert matrices CO$_2$ and H$_2$O. Particularly, in the H$_2$O ice there are alternative reaction pathways that potentially form products such as HOCH$_2$CN. We find that oxygen insertion into CH$_3$CN is a viable novel grain-surface pathway that can contribute to CH$_3$NCO and HOCH$_2$CN production and help explain the observational abundances.

The COMs CH$_3$CN, CH$_3$NCO, and HOCH$_2$CN are all attractive interstellar prebiotic molecules and have been detected in several regions of the ISM at different stages of star formation.  CH$_3$NCO can form peptide bonds while HOCH$_2$CN can form amino acid chains such as glycine and adenine \citep{pascal_prebiotic_2005, quenard_chemical_2018, rodriguez_nitrogen_2019, schwartz_acceleration_1982}. CH$_3$CN is the largest nitrile in protoplanetary disks \citep{bergner_survey_2018}. CH$_3$NCO has been detected in a variety of different environments, ranging from the surface of the comet 67P/Churyumov-Gerasimenko, to hot cores around high-mass protostars, to the molecular clouds of Sgr B2 N and Orion KL \citep{halfen_interslar_2015, vavra_millimeter_2022, martin-domenech_detection_2017, ligterink_alma-pils_2017, cernicharo_rigorous_2016, goesmann_organic_2015, gorai_identification_2021}. Although there are less detections for HOCH$_2$CN including a null detection in Sgr B(2N), it has been observed in both the inner hot corino and other cold envelope of a solar type protostar \citep{zeng_first_2019, zhao_glycolonitrile_2021, margules_submillimeter_2017}. CH$_3$NCO, and HOCH$_2$CN have been detected simultaneously in the low-mass protostar IRAS 16293B (although in different data sets) and the Class O intermediate mass protostar Serpens SMM1-a \citep{ligterink_prebiotic_2021}. 

Despite multiple detections, the formation pathways of CH$_3$NCO and HOCH$_2$CN in the ISM remain unclear. CH$_3$NCO is proposed to form via gas-phase methylation of HNCO or HOCN \citep{halfen_interslar_2015}, supported by its spatial correlation with HNCO in Orion \citep{cernicharo_rigorous_2016}. However, its absence in dark clouds and high abundance in Orion and Sgr B2 suggest a grain-surface origin \citep{ligterink_prebiotic_2021}. The radical-radical surface reaction CH$_3$ + OCN has been experimentally confirmed to proceed \citep{belloche_rotational_2017, ligterink_alma-pils_2017}, though the product yields are low due to competing channels such CH$_4$ + NCO formation. In summary, these models still fall short of matching observations in sources like IRAS 16293 and L1544 \citep{quenard_chemical_2018}.

HOCH$_2$CN has no known gas-phase route, but can form on grain surfaces via radical recombination of OH + CH$_2$CN or HOCH$_2$ + CN \citep{woon_formation_2020, bulak_photolysis_2021}. These proposed reactions remain untested. Experiments show HOCH$_2$CN can form from CN$^-$ + H$_2$CO in H$_2$O- or NH$_3$-rich ices acting as bases \citep{danger_hydroxyacetonitrile_2012, danger_formation_2014}. However, current gas-grain models such as UCLCHEM incorporating these reactions cannot reproduce the observed abundance in IRAS 16293B \citep{zeng_first_2019}. 

Aside from determining abundances, it is difficult to use observational data to establish a chemical link between -CN and -NCO molecules especially with limited detections of HOCH$_2$CN.  For example, in Serpens SMM1-a, HOCH$_2$CN and HNCO are enhanced, but CH$_3$NCO, CH$_3$CN, and NH$_2$CN are not \citep{ligterink_prebiotic_2021}. HOCH$_2$CN may instead require different reactants or physical conditions. However, abundance correlations do not necessarily imply formation pathways \citep{belloche_questioning_2020}. 

Such observational limitations reinforce the need for astrochemical surveys capable of probing whether laboratory identified pathways, such as the O($^1$D)-induced conversion of CH$_3$CN to CH$_3$NCO reported here, are active in interstellar environments. We need more unbiased detections across diverse sources particularly those where both HOCH$_2$CN and CH$_3$NCO are observed to compare their abundances not only relative to HNCO, but also to CH$_3$CN. Establishing these relationships will help constrain whether –CN and –NCO species share chemical ancestry or arise through distinct formation routes.

Overall, our work supports that additional grain-surface pathways can contribute to CH$_3$NCO and HOCH$_2$CN production and help explain the observational abundances. We propose an efficient pathway for CH$_3$NCO through \singlet atom insertion into CH$_3$CN on ice grains. In the presence of H$_2$O, we also demonstrate the feasibility of HOCH$_2$CN formation from CH$_3$CN and \singlet atoms. These are both novel grain pathways and distinct from thermal Strecker-like synthesis and radical recombination previously demonstrated for HOCH$_2$CN and CH$_3$NCO. In general, atoms have a lower barrier to diffusion compared to larger radical fragments, such that oxygen atoms still have enough mobility within the coldest ISM ices to encounter reactants and form more complex molecules even when radical recombination cannot occur. Building on previous work whereby O($^1$D) atoms reacted with CH$_4$ to form CH$_3$OH in cold ices \citep{bergner_methanol_2017}, we demonstrate that excited-state oxygen chemistry can also initiate CH$_3$NCO and HOCH$_2$CN formation from CH$_3$CN on grain surfaces. These pathway offers a plausible explanation for the presence of both molecules in protostellar sources such as IRAS 16293 and Serpens SMM1-a in the very early stages of star formation.

\section{Conclusion} \label{sec:Conclusion}
We investigated the UV irradiation of CH$_3$CN and O$_3$ ices to experimentally examine oxygen insertion. Our experiments demonstrate a highly efficient pathway to produce CH$_3$NCO and HOCH$_2$CN in astrophysical ice analogs. From our studies we conclude the following: 
\begin{enumerate}
\item UV irradiation of O$_3$ at 254 nm produces \singlet atoms that readily insert into CH$_3$CN at low temperatures to produce primarily CH$_3$NCO.

\item CH$_3$NCO formation is matrix-dependent, with likely lower yields observed in non-inert CO$_2$ and H$_2$O ices. The notably lower limit to the yield in H$_2$O suggests that alternative reaction pathways dominate, potentially leading to products such as HOCH$_2$CN.

\item We find no temperature dependence to O$_3$ destruction, CH$_3$CN destruction, or CH$_3$NCO formation rate from 10 to 40 K supporting that \singlet atom insertion is a barrierless process. 

\item We form CH$_3$NCO, although at a lower efficiency, from photolysis of O$_2$ with a H$_2$D$_2$ lamp at 160 nm due to simultaneous destruction of CH$_3$CN. Thus, CH$_3$NCO formation is not exclusive to O$_3$ and can proceed with different oxygen atom sources. 

\item Under cold ISM conditions, oxygen atom chemistry on ice grains can convert the nitrile CH$_3$CN into more complex organic molecules. We propose that this chemistry may help to explain detections of CH$_3$NCO and HOCH$_2$CN in cold sources during the very early stages of star formation.

\end{enumerate}

\noindent This work was supported by a grant from the Simons Foundation (686302, K.I.Ö.).





\clearpage
\appendix {} \label{appendix}
\vspace{-12pt}

\section{Methyl Cyanide Control Experiment} \label{appendix:blank}
\begin{figure*} [h]
\plotone{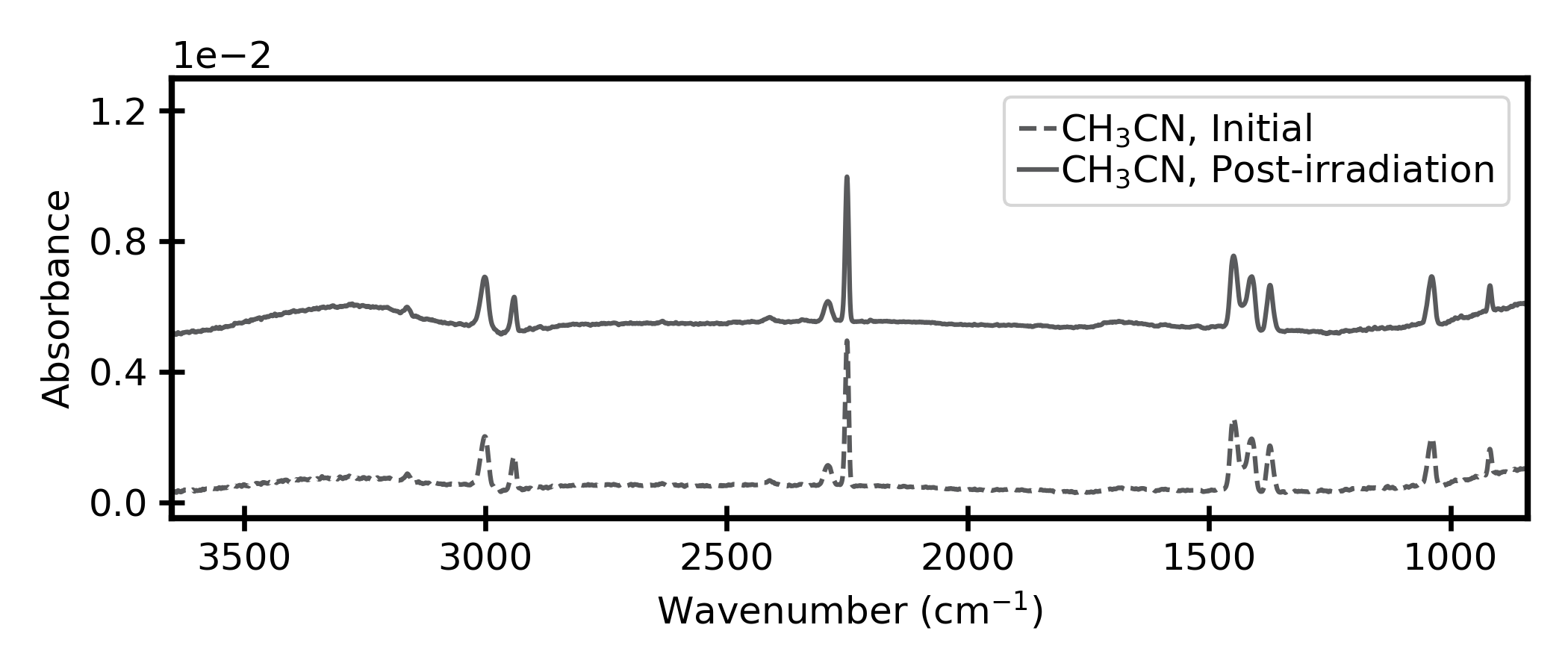}
\caption{Initial (dashed) and post-irradiation (solid) infrared spectra of CH$_3$CN at 40 K.
\label{fig:appendix_blank}}
\end{figure*}

Figure \ref{fig:appendix_blank} shows the initial (dashed) and post-irradiation (solid) infrared spectra before and after exposure to 254 nm UV at 40 K for CH$_3$CN. The absence of spectral changes confirms that CH$_3$CN is unreactive.

\section{Methyl Isocyanate Product Confirmation} \label{appendix:methyl isocyanate}

\begin{figure*} [h]
\plotone{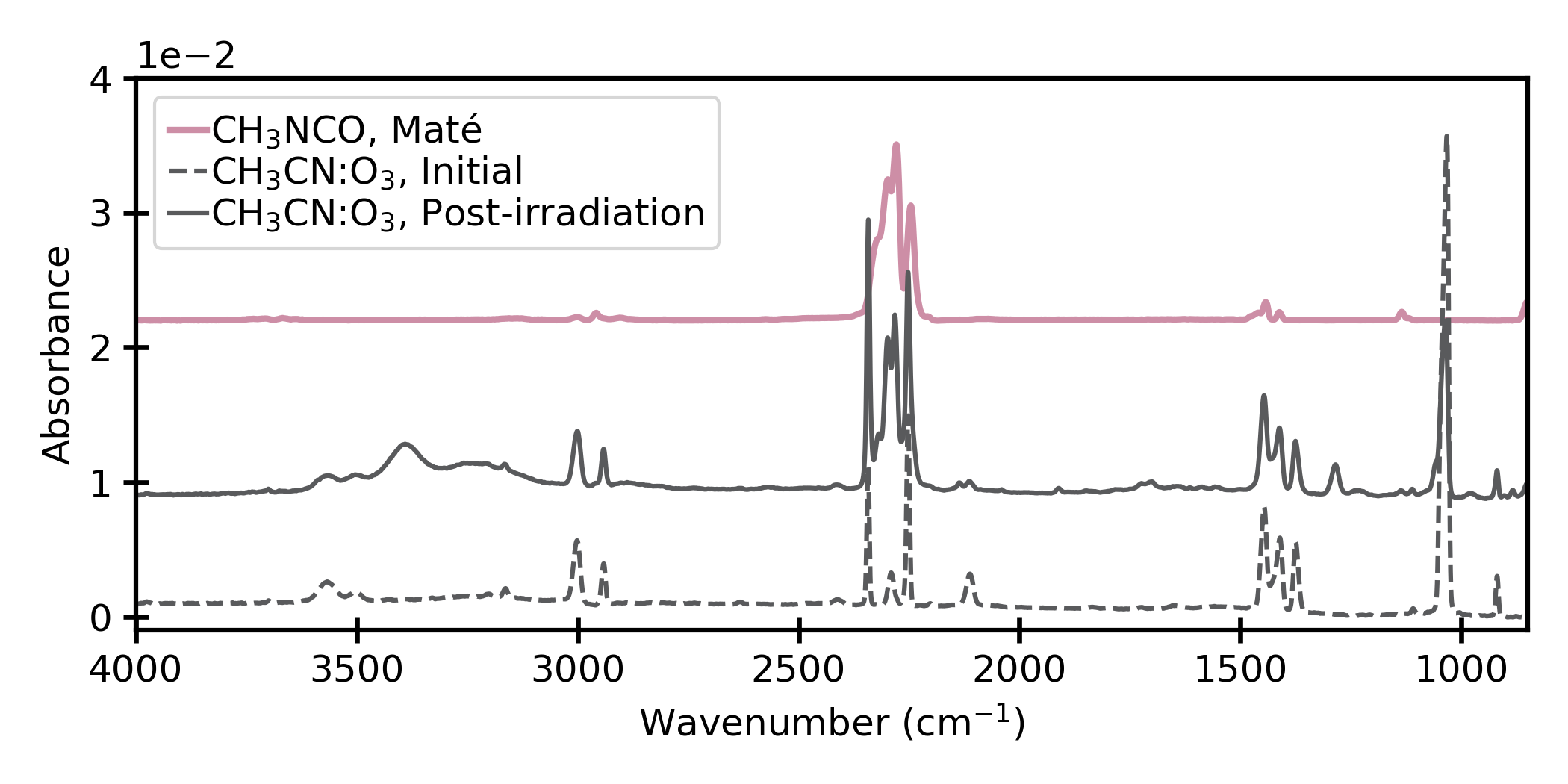}
\caption{Infrared spectra of a CH$_3$CN:O$_3$ ice before (dashed) and after (solid) exposure to 254 nm UV at 40 K overlaid with \citep{mate_laboratory_2017}.
\label{fig:appendix_mate}}
\end{figure*}

Figure \ref{fig:appendix_mate} shows the post-irradiation spectra for a CH$_3$CN:O$_3$ ice overlaid with the reference spectra from \citep{mate_laboratory_2017} for pure methyl isocyanate (CH$_3$NCO). We confirm the presence of CH$_3$NCO through features at 2320, 2297, 2282, and 2235 cm$^{-1}$, but there is also good agreement for the -CH$_3$ bending and stretching modes.

\newpage
\section{Rate Values} \label{appendix:curve_fitt}
Fitted kinetic parameters for O$_3$, CH$_3$CN, and CH$_3$NCO at 10 K, 20 K, and 40 K (Experiments $\#$2, 3, 4).

\begin{deluxetable}{llccc} [h]
\tablecaption{ \label{tab:appendix_constants}}
\tablewidth{0pt}
\tablehead{
\colhead{Molecule} & 
\colhead{Temp (K)} & 
\colhead{$k$ [cm$^2$ photon$^{-1}$]} & 
\colhead{$A_0$ or $SS$} & 
\colhead{$J$}
}
\startdata
O$_3$     & 10 & $(1.4 \pm 0.2) \times 10^{-17}$ & $0.71 \pm 0.04$ & $0.27 \pm 0.05$ \\
O$_3$     & 20 & $(1.6 \pm 0.1) \times 10^{-17}$ & $0.67 \pm 0.02$ & $0.32 \pm 0.02$ \\
O$_3$     & 40 & $(1.4 \pm 0.1) \times 10^{-17}$ & $0.71 \pm 0.03$ & $0.28 \pm 0.03$ \\
\hline
CH$_3$CN  & 10 & $(10 \pm 13) \times 10^{-18}$ & $0.35 \pm 0.02$ & $0.66 \pm 0.27$ \\
CH$_3$CN  & 20 & $(6.8 \pm 4.7) \times 10^{-18}$  & $0.51 \pm 0.23$ & $0.50 \pm 0.24$ \\
CH$_3$CN  & 40 & $(3.3 \pm 8.3) \times 10^{-18}$  & $1.0 \pm 2.1$ & $(1.0 \pm 0.6) \times 10^{-5}$ \\
\hline
CH$_3$NCO & 10 & $(1.4 \pm 0.1) \times 10^{-17}$  & $0.34 \pm 0.02$ & \nodata \\
CH$_3$NCO & 20 & $(1.6 \pm 0.2) \times 10^{-17}$  & $0.23 \pm 0.01$ & \nodata \\
CH$_3$NCO & 40 & $(1.5 \pm 0.2) \times 10^{-17}$  & $0.27 \pm 0.02$ & \nodata \\
\enddata
\end{deluxetable}

\clearpage
\section{Reactivity in Ice Matrices} 
\begin{figure*} [h]
\plotone{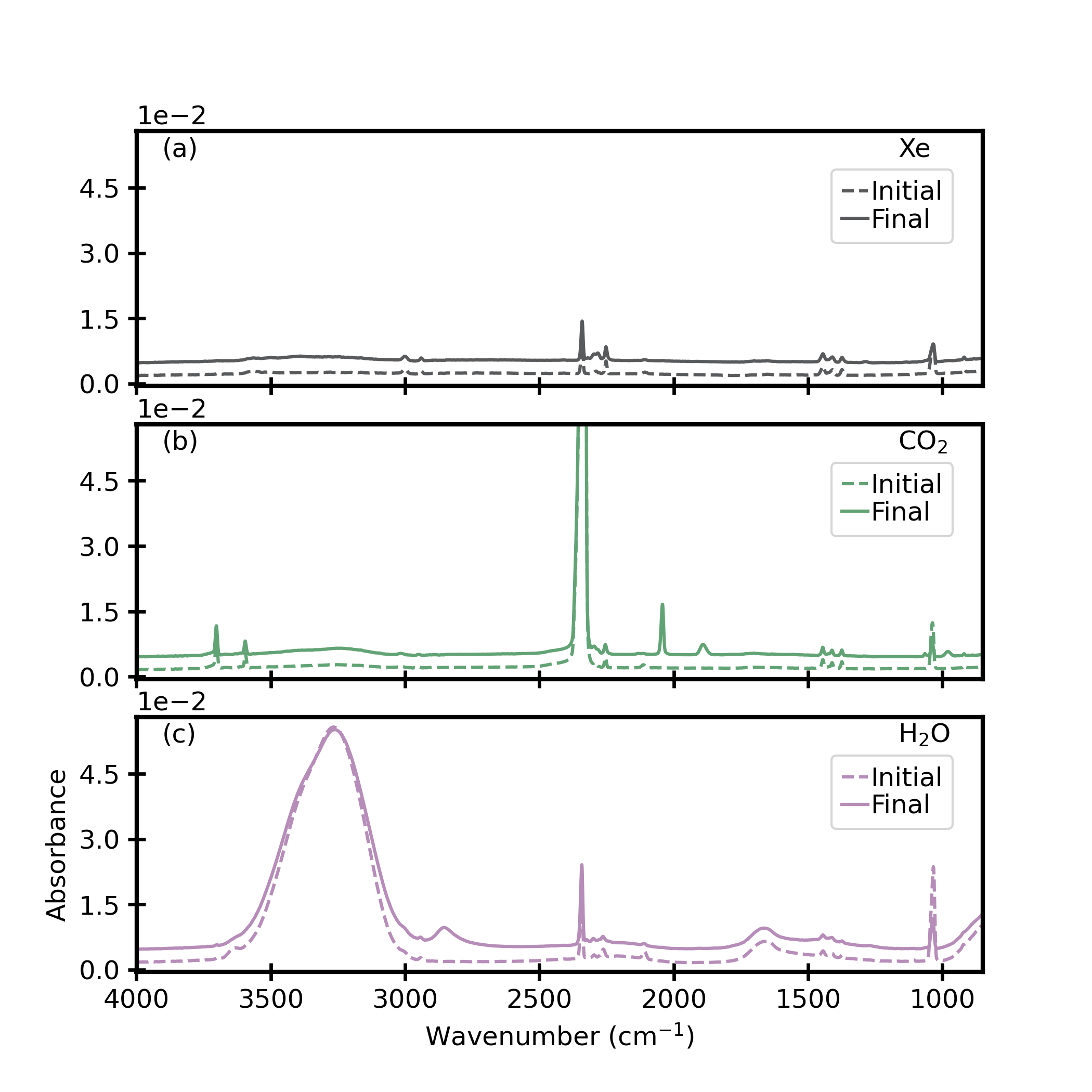}
\caption{Initial (dashed) and post-irradiation (solid) infrared spectra of CH$_3$CN:O$_3$ in a Xe (a), CO$_2$ (b), and H$_2$O (c) matrix at 40 K.
\label{fig:appendix_matrices}}
\end{figure*}

Figure \ref{fig:appendix_matrices} shows the initial (dashed) and post-irradiation (solid) infrared spectra before and after exposure to 254 nm UV at 40 K for CH$_3$CN:O$_3$ in Xe (a), CO$_2$ (b), and H$_2$O (c) ice matrices. While xenon is infrared inactive, the main CO$_2$ infrared feature is at 2340 cm$^{-1}$ and main H$_2$O feature is between 3600 and 3000 cm$^{-1}$. 

\clearpage
\bibliography{references_v2}{}

\begin{thebibliography}{}
\expandafter\ifx\csname natexlab\endcsname\relax\def\natexlab#1{#1}\fi
\providecommand{\url}[1]{\href{#1}{#1}}
\providecommand{\dodoi}[1]{doi:~\href{http://doi.org/#1}{\nolinkurl{#1}}}
\providecommand{\doeprint}[1]{\href{http://ascl.net/#1}{\nolinkurl{http://ascl.net/#1}}}
\providecommand{\doarXiv}[1]{\href{https://arxiv.org/abs/#1}{\nolinkurl{https://arxiv.org/abs/#1}}}

\bibitem[{Abdulgalil {et~al.}(2013)Abdulgalil, Marchione, Thrower, Collings, McCoustra, Islam, Palumbo, Congiu, \& Dulieu}]{abdulgalil_laboratory_2013}
Abdulgalil, A. G.~M., Marchione, D., Thrower, J.~D., {et~al.} 2013, Philosophical Transactions of the Royal Society A: Mathematical, Physical and Engineering Sciences, 371, \dodoi{10.1098/rsta.2011.0586}

\bibitem[{Altwegg {et~al.}(2019)Altwegg, Balsiger, \& Fuselier}]{altwegg_cometary_2019}
Altwegg, K., Balsiger, H., \& Fuselier, S.~A. 2019, Annual Review of Astronomy and Astrophysics, 57, 113, \dodoi{10.1146/annurev-astro-091918-104409}

\bibitem[{Altwegg {et~al.}(2017)Altwegg, Balsiger, Berthelier, Bieler, Calmonte, Fuselier, Goesmann, Gasc, Gombosi, Le~Roy, de~Keyser, Morse, Rubin, Schuhmann, Taylor, Tzou, \& Wright}]{altwegg_organics_2017}
Altwegg, K., Balsiger, H., Berthelier, J., {et~al.} 2017, Monthly Notices of the Royal Astronomical Society, 469, S130, \dodoi{10.1093/mnras/stx1415}

\bibitem[{Bar-nun {et~al.}(1985)Bar-nun, Herman, Laufer, \& Rappaport}]{bar-nun_trapping_1985}
Bar-nun, A., Herman, G., Laufer, D., \& Rappaport, M.~L. 1985, Icarus, 63, 317, \dodoi{10.1016/0019-1035(85)90048-X}

\bibitem[{Bar-Nun {et~al.}(2007)Bar-Nun, Notesco, \& Owen}]{bar-nun_trapping_2007}
Bar-Nun, A., Notesco, G., \& Owen, T. 2007, Icarus, 190, 655, \dodoi{10.1016/j.icarus.2007.03.021}

\bibitem[{Belloche {et~al.}(2009)Belloche, Garrod, Müller, Menten, Comito, \& Schilke}]{belloche_increased_2009}
Belloche, A., Garrod, R.~T., Müller, H. S.~P., {et~al.} 2009, Astronomy \& Astrophysics, 499, 215, \dodoi{10.1051/0004-6361/200811550}

\bibitem[{Belloche {et~al.}(2017)Belloche, Meshcheryakov, Garrod, Ilyushin, Alekseev, Motiyenko, Margulès, Müller, \& Menten}]{belloche_rotational_2017}
Belloche, A., Meshcheryakov, A.~A., Garrod, R.~T., {et~al.} 2017, Astronomy \& Astrophysics, 601, A49, \dodoi{10.1051/0004-6361/201629724}

\bibitem[{Belloche {et~al.}(2020)Belloche, Maury, Maret, Anderl, Bacmann, André, Bontemps, Cabrit, Codella, Gaudel, Gueth, Lefèvre, Lefloch, Podio, \& Testi}]{belloche_questioning_2020}
Belloche, A., Maury, A.~J., Maret, S., {et~al.} 2020, Astronomy \& Astrophysics, 635, A198, \dodoi{10.1051/0004-6361/201937352}

\bibitem[{Bergner {et~al.}(2018)Bergner, Guzmán, Öberg, Loomis, \& Pegues}]{bergner_survey_2018}
Bergner, J.~B., Guzmán, V.~G., Öberg, K.~I., Loomis, R.~A., \& Pegues, J. 2018, The Astrophysical Journal, 857, 69, \dodoi{10.3847/1538-4357/aab664}

\bibitem[{Bergner {et~al.}(2017)Bergner, Öberg, \& Rajappan}]{bergner_methanol_2017}
Bergner, J.~B., Öberg, K.~I., \& Rajappan, M. 2017, The Astrophysical Journal, 845, 29, \dodoi{10.3847/1538-4357/aa7d09}

\bibitem[{Bergner {et~al.}(2019)Bergner, Öberg, \& Rajappan}]{bergner_oxygen_2019}
---. 2019, The Astrophysical Journal, 874, 115, \dodoi{10.3847/1538-4357/ab07b2}

\bibitem[{Bondybey {et~al.}(1981)Bondybey, English, Mathews, \& Contolini}]{bondybey_cno_1981}
Bondybey, V.~E., English, J.~H., Mathews, C.~W., \& Contolini, R.~J. 1981, Chemical Physics Letters, 82, 208, \dodoi{10.1016/0009-2614(81)85140-8}

\bibitem[{Boogert {et~al.}(2015)Boogert, Gerakines, \& Whittet}]{boogert_observations_2015}
Boogert, A. C.~A., Gerakines, P.~A., \& Whittet, D. C.~B. 2015, Annual Review of Astronomy and Astrophysics, 53, 541, \dodoi{10.1146/annurev-astro-082214-122348}

\bibitem[{Bouilloud {et~al.}(2015)Bouilloud, Fray, Bénilan, Cottin, Gazeau, \& Jolly}]{bouilloud_bibliographic_2015}
Bouilloud, M., Fray, N., Bénilan, Y., {et~al.} 2015, Monthly Notices of the Royal Astronomical Society, 451, 2145, \dodoi{10.1093/mnras/stv1021}

\bibitem[{Brann {et~al.}(2020)Brann, Thompson, \& Sibener}]{brann_reaction_2020}
Brann, M.~R., Thompson, R.~S., \& Sibener, S.~J. 2020, The Journal of Physical Chemistry C, 124, 7205, \dodoi{10.1021/acs.jpcc.9b11439}

\bibitem[{Bulak {et~al.}(2021)Bulak, Paardekooper, Fedoseev, \& Linnartz}]{bulak_photolysis_2021}
Bulak, M., Paardekooper, D.~M., Fedoseev, G., \& Linnartz, H. 2021, Astronomy \& Astrophysics, 647, A82, \dodoi{10.1051/0004-6361/202039695}

\bibitem[{Callen {et~al.}(1990)Callen, Griffiths, Memmert, Harrington, Bushby, \& Norton}]{callen_adsorption_1990}
Callen, B.~W., Griffiths, K., Memmert, U., {et~al.} 1990, Surface Science, 230, 159, \dodoi{10.1016/0039-6028(90)90024-3}

\bibitem[{Canta {et~al.}(2023)Canta, Öberg, \& Rajappan}]{canta_formation_2023}
Canta, A., Öberg, K.~I., \& Rajappan, M. 2023, The Astrophysical Journal, 953, 81, \dodoi{10.3847/1538-4357/acda99}

\bibitem[{Carder {et~al.}(2021)Carder, Ochs, \& Herbst}]{carder_modelling_2021}
Carder, J.~T., Ochs, W., \& Herbst, E. 2021, Monthly Notices of the Royal Astronomical Society, 508, 1526, \dodoi{10.1093/mnras/stab2619}

\bibitem[{Carvalho \& Pilling(2020)}]{carvalho_photolysis_2020}
Carvalho, G.~A., \& Pilling, S. 2020, The Journal of Physical Chemistry A, 124, 8574, \dodoi{10.1021/acs.jpca.0c06229}

\bibitem[{Cernicharo {et~al.}(2016)Cernicharo, Kisiel, Tercero, Kolesniková, Medvedev, López, Fortman, Winnewisser, Lucia, Alonso, \& Guillemin}]{cernicharo_rigorous_2016}
Cernicharo, J., Kisiel, Z., Tercero, B., {et~al.} 2016, Astronomy \& Astrophysics, 587, L4, \dodoi{10.1051/0004-6361/201527531}

\bibitem[{Chaabouni {et~al.}(2000)Chaabouni, Schriver-Mazzuoli, \& Schriver}]{chaabouni_infrared_2000}
Chaabouni, H., Schriver-Mazzuoli, L., \& Schriver, A. 2000, The Journal of Physical Chemistry A, 104, 6962, \dodoi{10.1021/jp0008290}

\bibitem[{Collings {et~al.}(2004)Collings, Anderson, Chen, Dever, Viti, Williams, \& McCoustra}]{collings_laboratory_2004}
Collings, M.~P., Anderson, M.~A., Chen, R., {et~al.} 2004, Monthly Notices of the Royal Astronomical Society, 354, 1133, \dodoi{10.1111/j.1365-2966.2004.08272.x}

\bibitem[{Dalbouha {et~al.}(2016)Dalbouha, Senent, Komiha, \& Domínguez-Gómez}]{dalbouha_structural_2016}
Dalbouha, S., Senent, M.~L., Komiha, N., \& Domínguez-Gómez, R. 2016, The Journal of Chemical Physics, 145, 124309, \dodoi{10.1063/1.4963186}

\bibitem[{Danger {et~al.}(2012)Danger, Duvernay, Theulé, Borget, \& Chiavassa}]{danger_hydroxyacetonitrile_2012}
Danger, G., Duvernay, F., Theulé, P., Borget, F., \& Chiavassa, T. 2012, The Astrophysical Journal, 756, 11, \dodoi{10.1088/0004-637X/756/1/11}

\bibitem[{Danger {et~al.}(2014)Danger, Rimola, Mrad, Duvernay, Roussin, Theule, \& Chiavassa}]{danger_formation_2014}
Danger, G., Rimola, A., Mrad, N.~A., {et~al.} 2014, Physical Chemistry Chemical Physics, 16, 3360, \dodoi{10.1039/C3CP54034K}

\bibitem[{Daniely {et~al.}(2025)Daniely, Zamir, Eisenberg, Livshits, Piacentino, Bergner, Öberg, \& Stein}]{daniely_photochemical_2025}
Daniely, A., Zamir, A., Eisenberg, H.~R., {et~al.} 2025, The Journal of Chemical Physics, 162, 014303, \dodoi{10.1063/5.0214165}

\bibitem[{Dereka {et~al.}(2022)Dereka, Lewis, Keim, Snyder, \& Tokmakoff}]{dereka_characterization_2022}
Dereka, B., Lewis, N. H.~C., Keim, J.~H., Snyder, S.~A., \& Tokmakoff, A. 2022, The Journal of Physical Chemistry B, 126, 278, \dodoi{10.1021/acs.jpcb.1c09572}

\bibitem[{d'Hendecourt \& Allamandola(1986)}]{dhendecourt_time_1986}
d'Hendecourt, L.~B., \& Allamandola, L.~J. 1986, Astron. Astrophys. Suppl. Ser.; (France), 64:3.
\newblock \url{https://www.osti.gov/etdeweb/biblio/5322644}

\bibitem[{Goesmann {et~al.}(2015)Goesmann, Rosenbauer, Bredehöft, Cabane, Ehrenfreund, Gautier, Giri, Krüger, Le~Roy, MacDermott, McKenna-Lawlor, Meierhenrich, Caro, Raulin, Roll, Steele, Steininger, Sternberg, Szopa, Thiemann, \& Ulamec}]{goesmann_organic_2015}
Goesmann, F., Rosenbauer, H., Bredehöft, J.~H., {et~al.} 2015, Science, 349, aab0689, \dodoi{10.1126/science.aab0689}

\bibitem[{Gorai {et~al.}(2021)Gorai, Das, Shimonishi, Sahu, Mondal, Bhat, \& Chakrabarti}]{gorai_identification_2021}
Gorai, P., Das, A., Shimonishi, T., {et~al.} 2021, The Astrophysical Journal, 907, 108, \dodoi{10.3847/1538-4357/abc9c4}

\bibitem[{Halfen {et~al.}(2015)Halfen, Ilyushin, \& Ziurys}]{halfen_interslar_2015}
Halfen, D.~T., Ilyushin, V.~V., \& Ziurys, L.~M. 2015, The Astrophysical Journal Letters, 812, L5, \dodoi{10.1088/2041-8205/812/1/L5}

\bibitem[{Herbst \& van Dishoeck(2009)}]{herbst_complex_2009}
Herbst, E., \& van Dishoeck, E.~F. 2009, Annual Review of Astronomy and Astrophysics, 47, 427, \dodoi{10.1146/annurev-astro-082708-101654}

\bibitem[{Hickson \& Loison(2022)}]{hickson_kinetic_2022}
Hickson, K.~M., \& Loison, J.-C. 2022, The Journal of Physical Chemistry A, 126, 3903, \dodoi{10.1021/acs.jpca.2c01946}

\bibitem[{Hudson \& Moore(2004)}]{hudson_reactions_2004}
Hudson, R.~L., \& Moore, M.~H. 2004, Icarus, 172, 466, \dodoi{10.1016/j.icarus.2004.06.011}

\bibitem[{Hudson {et~al.}(2008)Hudson, Moore, Dworkin, Martin, \& Pozun}]{hudson_amino_2008}
Hudson, R.~L., Moore, M.~H., Dworkin, J.~P., Martin, M.~P., \& Pozun, Z.~D. 2008, Astrobiology, 8, 771, \dodoi{10.1089/ast.2007.0131}

\bibitem[{Ioppolo {et~al.}(2008)Ioppolo, Cuppen, Romanzin, van Dishoeck, \& Linnartz}]{ioppolo_laboratory_2008}
Ioppolo, S., Cuppen, H.~M., Romanzin, C., van Dishoeck, E.~F., \& Linnartz, H. 2008, The Astrophysical Journal, 686, 1474, \dodoi{10.1086/591506}

\bibitem[{Jones {et~al.}(2011)Jones, Bennett, \& Kaiser}]{jones_mechanistical_2011}
Jones, B.~M., Bennett, C.~J., \& Kaiser, R.~I. 2011, The Astrophysical Journal, 734, 78, \dodoi{10.1088/0004-637X/734/2/78}

\bibitem[{Kedzierski {et~al.}(2013)Kedzierski, Hein, Tiessen, Lukic, Trocchi, Mlinaric, \& McConkey}]{kedzierski_production_2013}
Kedzierski, W., Hein, J., Tiessen, C., {et~al.} 2013, Canadian Journal of Physics, 91, 1044, \dodoi{10.1139/cjp-2013-0255}

\bibitem[{Kim \& Kim(1992)}]{kim_fourier_1992}
Kim, H.~S., \& Kim, K. 1992, Bulletin of the Korean Chemical Society, 13, 520.
\newblock \url{https://koreascience.kr/article/JAKO199213464458311.page}

\bibitem[{Knoezinger {et~al.}(1993)Knoezinger, Beichert, Hermeling, \& Schrems}]{knoezinger_matrix_1993}
Knoezinger, E., Beichert, P., Hermeling, J., \& Schrems, O. 1993, The Journal of Physical Chemistry, 97, 1324, \dodoi{10.1021/j100109a013}

\bibitem[{Lauck {et~al.}(2015)Lauck, Karssemeijer, Shulenberger, Rajappan, Öberg, \& Cuppen}]{lauck_co_2015}
Lauck, T., Karssemeijer, L., Shulenberger, K., {et~al.} 2015, The Astrophysical Journal, 801, 118, \dodoi{10.1088/0004-637X/801/2/118}

\bibitem[{Lee {et~al.}(1977)Lee, Slanger, Black, \& Sharpless}]{lee_quantum_1977}
Lee, L.~C., Slanger, T.~G., Black, G., \& Sharpless, R.~L. 1977, The Journal of Chemical Physics, 67, 5602, \dodoi{10.1063/1.434759}

\bibitem[{Ligterink {et~al.}(2017)Ligterink, Coutens, Kofman, Müller, Garrod, Calcutt, Wampfler, Jørgensen, Linnartz, \& van Dishoeck}]{ligterink_alma-pils_2017}
Ligterink, N. F.~W., Coutens, A., Kofman, V., {et~al.} 2017, Monthly Notices of the Royal Astronomical Society, 469, 2219, \dodoi{10.1093/mnras/stx890}

\bibitem[{Ligterink {et~al.}(2021)Ligterink, Ahmadi, Coutens, Tychoniec, Calcutt, Dishoeck, Linnartz, Jørgensen, Garrod, \& Bouwman}]{ligterink_prebiotic_2021}
Ligterink, N. F.~W., Ahmadi, A., Coutens, A., {et~al.} 2021, Astronomy \& Astrophysics, 647, A87, \dodoi{10.1051/0004-6361/202039619}

\bibitem[{Loeffler {et~al.}(2006)Loeffler, Teolis, \& Baragiola}]{loeffler_decomposition_2006}
Loeffler, M.~J., Teolis, B.~D., \& Baragiola, R.~A. 2006, The Journal of Chemical Physics, 124, 104702, \dodoi{10.1063/1.2171967}

\bibitem[{Loomis {et~al.}(2018)Loomis, Cleeves, Öberg, Aikawa, Bergner, Furuya, Guzman, \& Walsh}]{loomis_distribution_2018}
Loomis, R.~A., Cleeves, L.~I., Öberg, K.~I., {et~al.} 2018, The Astrophysical Journal, 859, 131, \dodoi{10.3847/1538-4357/aac169}

\bibitem[{Margulès {et~al.}(2017)Margulès, McGuire, Senent, Motiyenko, Remijan, \& Guillemin}]{margules_submillimeter_2017}
Margulès, L., McGuire, B.~A., Senent, M.~L., {et~al.} 2017, Astronomy \& Astrophysics, 601, A50, \dodoi{10.1051/0004-6361/201628551}

\bibitem[{Marshall \& Burkholder(2024)}]{marshall_kinetics_2024}
Marshall, P., \& Burkholder, J.~B. 2024, ACS Earth and Space Chemistry, 8, 1933, \dodoi{10.1021/acsearthspacechem.4c00176}

\bibitem[{Martín-Doménech {et~al.}(2017)Martín-Doménech, Rivilla, Jiménez-Serra, Quénard, Testi, \& Martín-Pintado}]{martin-domenech_detection_2017}
Martín-Doménech, R., Rivilla, V.~M., Jiménez-Serra, I., {et~al.} 2017, Monthly Notices of the Royal Astronomical Society, 469, 2230, \dodoi{10.1093/mnras/stx915}

\bibitem[{Martín-Doménech {et~al.}(2020)Martín-Doménech, Öberg, \& Rajappan}]{martin-domenech_formation_2020}
Martín-Doménech, R., Öberg, K.~I., \& Rajappan, M. 2020, The Astrophysical Journal, 894, 98, \dodoi{10.3847/1538-4357/ab84e8}

\bibitem[{Matsumi \& Kawasaki(2003)}]{matsumi_photolysis_2003}
Matsumi, Y., \& Kawasaki, M. 2003, Chemical Reviews, 103, 4767, \dodoi{10.1021/cr0205255}

\bibitem[{Maté {et~al.}(2018)Maté, Molpeceres, Tanarro, Peláez, Guillemin, Cernicharo, \& Herrero}]{mate_stability_2018}
Maté, B., Molpeceres, G., Tanarro, I., {et~al.} 2018, The Astrophysical Journal, 861, 61, \dodoi{10.3847/1538-4357/aac826}

\bibitem[{Maté {et~al.}(2017)Maté, Molpeceres, Timón, Tanarro, Escribano, Guillemin, Cernicharo, \& Herrero}]{mate_laboratory_2017}
Maté, B., Molpeceres, G., Timón, V., {et~al.} 2017, Monthly Notices of the Royal Astronomical Society, 470, 4222, \dodoi{10.1093/mnras/stx1461}

\bibitem[{McGuire(2022)}]{mcguire_2021_2022}
McGuire, B.~A. 2022, The Astrophysical Journal Supplement Series, 259, 30, \dodoi{10.3847/1538-4365/ac2a48}

\bibitem[{Mielke {et~al.}(1989)Mielke, Hawkins, \& Andrews}]{mielke_matrix_1989}
Mielke, Z., Hawkins, M., \& Andrews, L. 1989, The Journal of Physical Chemistry, 93, 558, \dodoi{10.1021/j100339a015}

\bibitem[{Mifsud {et~al.}(2022)Mifsud, Kaňuchová, Ioppolo, Herczku, Traspas~Muiña, Field, Hailey, Juhász, Kovács, Mason, McCullough, Pavithraa, Rahul, Paripás, Sulik, Chou, Lo, Das, Cheng, Rajasekhar, Bhardwaj, \& Sivaraman}]{mifsud_mid-ir_2022}
Mifsud, D.~V., Kaňuchová, Z., Ioppolo, S., {et~al.} 2022, Journal of Molecular Spectroscopy, 385, 111599, \dodoi{10.1016/j.jms.2022.111599}

\bibitem[{Moll {et~al.}(1966)Moll, Clutter, \& Thompson}]{moll_carbon_1966}
Moll, N.~G., Clutter, D.~R., \& Thompson, W.~E. 1966, The Journal of Chemical Physics, 45, 4469, \dodoi{10.1063/1.1727526}

\bibitem[{Pascal {et~al.}(2005)Pascal, Boiteau, \& Commeyras}]{pascal_prebiotic_2005}
Pascal, R., Boiteau, L., \& Commeyras, A. 2005, in Prebiotic {Chemistry}, ed. P.~Walde (Berlin, Heidelberg: Springer), 69--122, \dodoi{10.1007/b136707}

\bibitem[{Pasinszki \& Westwood(2001)}]{pasinszki_gas-phase_2001}
Pasinszki, T., \& Westwood, N. P.~C. 2001, The Journal of Physical Chemistry A, 105, 1244, \dodoi{10.1021/jp002851z}

\bibitem[{Quénard {et~al.}(2018)Quénard, Jiménez-Serra, Viti, Holdship, \& Coutens}]{quenard_chemical_2018}
Quénard, D., Jiménez-Serra, I., Viti, S., Holdship, J., \& Coutens, A. 2018, Monthly Notices of the Royal Astronomical Society, 474, 2796, \dodoi{10.1093/mnras/stx2960}

\bibitem[{Rachid {et~al.}(2022)Rachid, Rocha, \& Linnartz}]{rachid_infrared_2022}
Rachid, M.~G., Rocha, W. R.~M., \& Linnartz, H. 2022, Astronomy \& Astrophysics, 665, A89, \dodoi{10.1051/0004-6361/202243417}

\bibitem[{Rodriguez {et~al.}(2019)Rodriguez, House, Smith, Roberts, \& Callahan}]{rodriguez_nitrogen_2019}
Rodriguez, L.~E., House, C.~H., Smith, K.~E., Roberts, M.~R., \& Callahan, M.~P. 2019, Scientific Reports, 9, 9281, \dodoi{10.1038/s41598-019-45310-z}

\bibitem[{Scheltinga {et~al.}(2018)Scheltinga, Ligterink, Boogert, Dishoeck, \& Linnartz}]{scheltinga_infrared_2018}
Scheltinga, J. T.~v., Ligterink, N. F.~W., Boogert, A. C.~A., Dishoeck, E. F.~v., \& Linnartz, H. 2018, Astronomy \& Astrophysics, 611, A35, \dodoi{10.1051/0004-6361/201731998}

\bibitem[{Schwartz \& Goverde(1982)}]{schwartz_acceleration_1982}
Schwartz, A.~W., \& Goverde, M. 1982, Journal of Molecular Evolution, 18, 351, \dodoi{10.1007/BF01733902}

\bibitem[{Simon {et~al.}(2019)Simon, Öberg, Rajappan, \& Maksiutenko}]{simon_entrapment_2019}
Simon, A., Öberg, K.~I., Rajappan, M., \& Maksiutenko, P. 2019, The Astrophysical Journal, 883, 21, \dodoi{10.3847/1538-4357/ab32e5}

\bibitem[{Sivaraman {et~al.}(2007)Sivaraman, Jamieson, Mason, \& Kaiser}]{sivaraman_temperature-dependent_2007}
Sivaraman, B., Jamieson, C.~S., Mason, N.~J., \& Kaiser, R.~I. 2007, The Astrophysical Journal, 669, 1414, \dodoi{10.1086/521216}

\bibitem[{Sun {et~al.}(2010)Sun, Tang, Jia, Wang, Sun, Feng, Pan, Hao, \& Wang}]{sun_theoretical_2010}
Sun, J., Tang, Y., Jia, X., {et~al.} 2010, The Journal of Chemical Physics, 132, 064301, \dodoi{10.1063/1.3292570}

\bibitem[{Ung(1974)}]{ung_photolysis_1974}
Ung, A. Y.~M. 1974, Chemical Physics Letters, 28, 603, \dodoi{10.1016/0009-2614(74)80117-X}

\bibitem[{Vanuzzo {et~al.}(2016)Vanuzzo, Balucani, Leonori, Stranges, Nevrly, Falcinelli, Bergeat, Casavecchia, \& Cavallotti}]{vanuzzo_reaction_2016}
Vanuzzo, G., Balucani, N., Leonori, F., {et~al.} 2016, The Journal of Physical Chemistry A, 120, 4603, \dodoi{10.1021/acs.jpca.6b01563}

\bibitem[{Vávra {et~al.}(2022)Vávra, Kolesniková, Belloche, Garrod, Koucký, Uhlíková, Luková, Guillemin, Kania, Müller, Menten, \& Urban}]{vavra_millimeter_2022}
Vávra, K., Kolesniková, L., Belloche, A., {et~al.} 2022, Astronomy \& Astrophysics, 666, A50, \dodoi{10.1051/0004-6361/202243627}

\bibitem[{Walsh {et~al.}(2014)Walsh, Millar, Nomura, Herbst, Weaver, Aikawa, Laas, \& Vasyunin}]{walsh_complex_2014}
Walsh, C., Millar, T.~J., Nomura, H., {et~al.} 2014, A\&A, 563, \dodoi{10.1051/0004-6361/201322446}

\bibitem[{Woon(2020)}]{woon_formation_2020}
Woon, D.~E. 2020, The Astrophysical Journal, 906, 20, \dodoi{10.3847/1538-4357/abc691}

\bibitem[{Zeng {et~al.}(2019)Zeng, Quénard, Jiménez-Serra, Martín-Pintado, Rivilla, Testi, \& Martín-Doménech}]{zeng_first_2019}
Zeng, S., Quénard, D., Jiménez-Serra, I., {et~al.} 2019, Monthly Notices of the Royal Astronomical Society: Letters, 484, L43, \dodoi{10.1093/mnrasl/slz002}

\bibitem[{Zhao {et~al.}(2021)Zhao, Quan, Zhang, Feng, Zhou, Li, Meng, Chang, Yang, He, \& Ma}]{zhao_glycolonitrile_2021}
Zhao, G., Quan, D., Zhang, X., {et~al.} 2021, The Astrophysical Journal Supplement Series, 257, 26, \dodoi{10.3847/1538-4365/ac17ee}

\bibitem[{Zhu \& Gordon(1990)}]{zhu_production_1990}
Zhu, Y., \& Gordon, R.~J. 1990, The Journal of Chemical Physics, 92, 2897, \dodoi{10.1063/1.457937}

\bibitem[{Öberg(2016)}]{oberg_photochemistry_2016}
Öberg, K.~I. 2016, Chemical Reviews, 116, 9631, \dodoi{10.1021/acs.chemrev.5b00694}

\bibitem[{Öberg {et~al.}(2015)Öberg, Guzmán, Furuya, Qi, Aikawa, Andrews, Loomis, \& Wilner}]{oberg_comet-like_2015}
Öberg, K.~I., Guzmán, V.~V., Furuya, K., {et~al.} 2015, Nature, 520, 198, \dodoi{10.1038/nature14276}

\end{thebibliography}
\bibliographystyle{aasjournal}

\end{document}